\documentclass[12pt, 
	oneside,		  
	a4paper,			
	english,			
	french,		        
	spanish,			
	brazil				
	]{abntex2}
	
\usepackage{lmodern}			    
\usepackage[T1]{fontenc}		  
\usepackage[utf8]{inputenc}		
\usepackage{lastpage}			    
\usepackage{indentfirst}		  
\usepackage{color}				    
\usepackage{graphicx}			    
\usepackage{subfig}
\usepackage{microtype} 		  	

\usepackage{amsmath,amsthm,amssymb}   
\usepackage{indentfirst} 
\usepackage[portuguese, ruled, linesnumbered,commentsnumbered, algo2e, vlined, lined, boxed, algochapter]{algorithm2e} 

\usepackage{algorithm}
\usepackage{algorithmic}

\usepackage{hyperref}
\usepackage[brazilian,hyperpageref]{backref}	 
\usepackage[alf]{abntex2cite}	
\usepackage{etoolbox}

\usepackage[font={bf, small}, labelsep=endash, labelfont=bf]{caption}	

\usepackage{multirow}
\usepackage[table,xcdraw]{xcolor}
\usepackage{float}

\usepackage{lipsum} 

\usepackage{rotating}
\usepackage{tikz}

\usepackage{pdfpages}

\usepackage{listings}
\usepackage{color}
\definecolor{lightgray2}{HTML}{F3F3F3}
\definecolor{lightgray}{rgb}{.9,.9,.9}
\definecolor{darkgray}{rgb}{.4,.4,.4}
\definecolor{darkred}{RGB}{139,0,0}
\definecolor{purple}{rgb}{0.65, 0.12, 0.82}
\definecolor{darkgreen}{RGB}{0,100,0}
\definecolor{green}{RGB}{0,128,0}
\definecolor{vividblue}{rgb}{0, 0, 255}

\lstdefinelanguage{JavaScript}{
  keywords={typeof, new, true, false, catch, function, return, null, catch, switch, var, let, const, require, async, if, in, while, do, else, case, break, interface, public, private},
  keywordstyle=\color{vividblue}\bfseries,
  ndkeywords={class, export, boolean, throw, implements, import, this, as, from, await, User, constructor, Array, any, number, string, integer},
  ndkeywordstyle=\color{darkgray}\bfseries,
  identifierstyle=\color{black},
  sensitive=false,
  comment=[l]{//},
  morecomment=[s]{/*}{*/},
  commentstyle=\color{green}\ttfamily,
  stringstyle=\color{red}\ttfamily,
  morestring=[b]',
  morestring=[b]"
}

\usepackage[export]{adjustbox}

\usepackage{tikz}
\usetikzlibrary{arrows.meta}
\tikzset{%
  >={Latex[width=2mm,length=2mm]},
            base/.style = {rectangle, rounded corners, draw=black,
                           minimum width=4cm, minimum height=1cm,
                           text centered, font=\sffamily},
  activityStarts/.style = {base, fill=blue!30},
       startstop/.style = {base, fill=red!30},
    activityRuns/.style = {base, fill=green!30},
         process/.style = {base, minimum width=2.5cm, fill=orange!15,
                           font=\ttfamily},
    decision/.style = {diamond, fill=pink!30},
}

\usepackage{upgreek}
\usepackage{fourier}
\usepackage{erewhon}
\usepackage{gensymb}
\usepackage{tablefootnote}
\usepackage[frozencache=true,cachedir=minted-cache]{minted} 
\makeatother

\definecolor{blue}{RGB}{25,25,112}

\makeatletter 
\hypersetup{ 
	pdftitle={\@title}, 
	pdfauthor={\@author},
    pdfsubject={\imprimirpreambulo},
	pdfcreator={LaTeX with abnTeX2},
	pdfkeywords={abnt}{latex}{abntex}{abntex2}{trabalho acadêmico}, 
	colorlinks=true,     
    linkcolor=black,          	
    citecolor=black,        		
    filecolor=magenta,      	
    urlcolor=blue,
	bookmarksdepth=4 }
\makeatother

\usepackage{tgtermes}
\renewcommand{\ABNTEXchapterfont}{\rmfamily\bfseries}

\providecommand{\imprimirsigla}{}
\newcommand{\sigla}[1]{\renewcommand{\imprimirsigla}{#1}}
\providecommand{\imprimiruf}{}
\newcommand{\uf}[1]{\renewcommand{\imprimiruf}{#1}}
\providecommand{\imprimircurso}{}
\newcommand{\curso}[1]{\renewcommand{\imprimircurso}{#1}}
\providecommand{\imprimirinstituto}{}
\newcommand{\instituto}[1]{\renewcommand{\imprimirinstituto}{#1}}
\providecommand{\imprimirdepartamento}{}
\newcommand{\departamento}[1]{\renewcommand{\imprimirdepartamento}{#1}}
\providecommand{\imprimirano}{}
\newcommand{\ano}[1]{\renewcommand{\imprimirano}{#1}}
\providecommand{\imprimirgrau}{}
\newcommand{\grau}[1]{\renewcommand{\imprimirgrau}{#1}}

\renewcommand{\imprimircapa}{  
\begin{capa}
\hspace{-1cm} \begin{minipage}[b]{0.15\linewidth}
\end{minipage} 
\hspace{-1cm} \begin{minipage}[b]{\linewidth}
              \begin{center}{\large
                      {\textsc{\imprimirinstituicao}}~-~{\textsc{\imprimirinstituto}} 
                      \break
                  }\end{center} 
              \end{minipage} 
\begin{minipage}[b]{0.15\linewidth}
\end{minipage}
\vfill
        \begin{center}
		{\textsc {\large \textbf{\imprimirautor}}}
		\\
		\vfill
        {\textsc{\huge \textbf{\imprimirtitulo}}}  \\
				\vfill
				{\textsc {\large \imprimirorientador}} 	
				\vfill
        {\large{\imprimirlocal~- \imprimiruf \\ \imprimirano  }}
        \end{center}
\end{capa}   } 

\renewcommand{\imprimirfolhaderosto}{
    \begin{center}
    {\textsc {\large \imprimirautor}}  \\
		\vfill
		{\textsc{\Huge \textbf{\imprimirtitulo}}}
    \end{center}
    \vfill 
    \begin{flushright} 
    \parbox{0.6\linewidth}{
		\imprimirtipotrabalho~ referente ao trabalho de conclusão de curso apresentado ao Curso de \imprimircurso~do \imprimirinstituicao, ~Campus Passo Fundo, como requisito parcial para a obtenção do título de \imprimirgrau. \\
		\vfill
		\textbf{\imprimirorientadorRotulo}~\imprimirorientador \\
		\vfill 
		\textbf{\imprimircoorientadorRotulo}~\imprimircoorientador}
   \end{flushright} 
   
	 \vfill
   \begin{center}
   {\large{\imprimirlocal~- \imprimiruf \\ \imprimirdata}}
   \end{center} }  

\titulo{Projeto e Implementação de uma DApp para armazenar dados de saúde \\
        \vspace{2cm}
        Design and Implementation of a DApp to store health data}
\autor{Christofer L. Sega \and Anubis G. de M. Rossetto \and 
Valderi R. Q. Leithardt } 
\local{Passo Fundo} \uf{RS}
\data{2022, v1} \ano{2022, v1}
\instituicao{Instituto Federal de Educação, Ciência e Tecnologia Sul-rio-grandense} \sigla{IFSul}
\instituto{Campus Passo Fundo}
\departamento{Departamento de Informática}
\curso{Ciências da Computação}	
\tipotrabalho{Relatório Técnico}
\grau{Bacharel em Ciências da Computação}

\makeindex   

\begin{document} 

\frenchspacing  

\pagenumbering{roman}

\imprimircapa  
\imprimirfolhaderosto 
\begin{agradecimentos}

A conquista de finalizar esse trabalho, só foi possível pelo conjunto dos esforços de várias pessoas, assim agradeço:

A minha mãe Carla e meu pai Claudio por estarem sempre dispostos a fazerem tudo dentro do alcance para me ajudar e fornecer todo o apoio necessário para continuar a seguir em frente, mesmo nos momentos mais difíceis.

A minha avó Sulmi, tia Rachel e prima Amanda por estarem presentes me apoiando e por me ajudarem com toda a questão de mudança de cidade para realizar o curso.

Ao meu irmão Junior, que mesmo longe, me apoiou da forma que conseguia.

A minha orientadora Anubis que durante todo o processo me guiou, instigou e me motivou para a elaboração deste trabalho, possibilitando que  conseguíssemos resultados incríveis.

Ao meu coorientador Valderi que mesmo distante forneceu o auxílio necessário para a construção do trabalho.

Aos meus professores do curso de Ciência da Computação que através de seus ensinamentos e companheirismo, permitiram que eu conseguisse finalizar este ciclo, em especial ao professor André pela parceria.

Aos meus amigos que fiz durante o trajeto, principalmente ao Bruno, Eliel e William, sempre apoiando um ao outro para continuar firme sem desistir da jornada.

Enfim, quero agradecer a todas as pessoas que ajudaram de forma direta e indireta na conquista de finalizar esse trabalho.

\end{agradecimentos}
\setlength{\absparsep}{18pt} 
\begin{resumo}
Este trabalho apresenta o projeto e implementação de uma aplicação descentralizada (DApp) que visa  garantir a privacidade dos dados relacionados à área da saúde, que são armazenados e compartilhados dentro de uma rede blockchain. Para tanto é empregado o uso de criptografia com os algoritmos RSA, ECC e AES. São apresentadas as plataformas, as tecnologias, as ferramentas e as bibliotecas necessárias para o desenvolvimento, bem como detalhes da implementação.

 \vspace{\onelineskip}
 \noindent
 \textbf{Palavras-chave}: Blockchain. Criptografia. Dados da saúde; DApp. Privacidade.
\end{resumo}

\begin{resumo}[Abstract]
 \begin{otherlanguage*}{english}
This work presents the design and implementation of a decentralized application (DApp) that aims to guarantee the privacy of data related to the health area, which are stored and shared within a blockchain network.  For this, encryption with RSA, ECC and AES algorithms is used. The platforms, technologies, tools and libraries required for development are presented, as well as implementation details.

\vspace{\onelineskip}
\end{otherlanguage*}
\end{resumo}

\pdfbookmark[0]{\listfigurename}{lof}
\listoffigures*   
\cleardoublepage

\begin{siglas}
    \item[ABI] Application Binary Interface
    \item[API] Application Programming Interface
    \item[CID] Content Identifier
    \item[DAG] Directed Acyclic Graph
	\item[DApp] Decentralized Application
	\item[DDoS] Distributed Denial of Service
	\item[DHT] Distributed Hash Table
	\item[DNS] Domain Name System
	\item[DOM] Document Object Model
	\item[ECC] Elliptic Curve Cryptography
	\item[ECIES] Elliptic Curve Integrated Encryption Scheme
	\item[ETH] Ether 
	\item[EVM] Ethereum Virtual Machine
	\item[HTTPS] Hypertext Transfer Protocol Secure 
	\item[IPFS] InterPlanetary File System
	\item[IPNS] InterPlanetary Name System
	\item[JSON] JavaScript Object Notation
	\item[JSX] JavaScript XML
	\item[POW] Proof-of-Work
	\item[RPC] Remote Procedure Calling
	\item[RSA] Rivest-Shamir-Adleman
	\item [WSS] Windows Sharepoint Services
\end{siglas}

\pdfbookmark[0]{\contentsname}{toc}
\tableofcontents*
\cleardoublepage

\pagenumbering{arabic} \setcounter{page}{1}
\textual 
\chapter{Introdução} \label{Introducao}

 A Blockchain proposta inicialmente por \citeonline{nakamoto2008bitcoin}, 
    permite que os dados fiquem publicamente visíveis para todos na rede blockchain, 
    consequentemente é importante que essas informações sejam 
    criptografadas antes de serem armazenadas. Desta forma, é possível garantir 
    a confidencialidade dos dados, pois manter o conteúdo da
    transação privado ajudará a reduzir o risco de vinculação do 
    pseudônimo a identidade real do usuário da Blockchain, o que é 
    fundamental para promover o compartilhamento baseado na 
    necessidade de saber \cite{ZhangSecurityBlock}.
    
    Este trabalho traz o detalhamento do projeto e da implementação de uma arquitetura que foi proposta para garantir a privacidade dos dados relacionados à área da saúde que são armazenados dentro de uma rede blockchain e no IPFS, de maneira descentralizada, através do uso de criptografia com os algoritmos Rivest-Shamir-Adleman (RSA), ECC e Advanced Encryption Standard (AES).
    
    A arquitetura prevê seis componentes que são: usuário, Dapp, MetaMask, Blockchain, IPFS e criptografia. O objetivo deste trabalho é trazer os fundamentos para a proposta da arquitetura, as ferramentas que foram utilizadas na  implementação, bem como detalhar como foram realizadas as comunicações entre estes componentes.  Cabe ressaltar que os algoritmos RSA e ECC tem a mesma função, assim, foram desenvolvidas bibliotecas de solução com ambos os algoritmos para poder conduzir uma avaliação sobre o impacto destes na arquitetura. Já o algoritmo AES é utilizado para fazer a criptografia do arquivo a ser enviado para o IPFS. Portanto este documentos traz um maior detalhamento quanto a implmentação e demonstra  protótipos da interface desenvolvida para interação do usuário.
    
    Este documento está organizado da seguinte forma: a primeira parte  aborda os fundamentos das tecnologias empregadas, bem como as ferramentas utilizadas. A segunda parte  apresenta a arquitetura proposta. Na terceira parte traz um detalhamento quanto  ao desenvolvimento da aplicação. Por fim, são apresentadas as conclusões obtidas a partir da proposta e implementação da arquitetura, além da realização de um apontamento para os próximos passos a serem realizados.


 
\chapter{Fundamentos}

\section{Blockchain}
	
    A blockchain foi originalmente introduzida, ou teve maior reconhecimento, quando Nakamoto em seu trabalho propôs um sistema financeiro utilizando a blockchain para registrar todas as transferências da moeda digital Bitcoin de forma segura e confiável \cite{FENG201945}. 
    
    Essa tecnologia é como um livro-razão descentralizado, distribuído e imutável, formado por uma coleção de registros que são criptograficamente vinculados, esta coleção é mais conhecida como corrente de blocos que armazenam transações ou eventos \cite{HEWA2021102857}. Este livro-razão é compartilhado com todos os membros participantes (nós) da rede blockchain.
    
    As transações que são realizadas entre os membros de uma blockchain devem ser aprovadas pelos nós mineradores antes de serem confirmadas e adicionadas a rede blockchain. Dessa forma, para iniciar o processo de mineração, a transação é transmitida a todos os nós da rede e os nós que são mineradores vão organizar as transações em um bloco, verificar as transações no bloco e transmitir o bloco e a sua verificação usando um protocolo de consenso, por exemplo o Proof of Work (POW), para obter a aprovação da rede \cite{ZhangSecurityBlock}.  Quando os demais nós verificarem se todas as transações contidas no bloco são válidas, o bloco pode ser adicionado a blockchain através de uma função hash criptográfica que conecta os blocos da estrutura, onde o hash do bloco \emph{n} está vinculado ao hash do bloco \emph{n+1} \cite{inbook}.
    
    Dentre as características da blockchain, as mais importantes segundo \citeonline{HEWA2021102857} são:
    
    \begin{itemize}    
       \item Descentralização: concede autoridade para os membros da rede, garantindo redundância em contraste com os sistemas centralizados operados por um terceiro confiável. A descentralização reduz o risco de falhas e acaba melhorando a confiança do serviço com disponibilidade garantida;
       
       \item Imutabilidade: os registros de transações no livro-razão, distribuídos entre os nós, são permanentes e inalteráveis. A imutabilidade é uma característica que difere dos sistemas de banco de dados centralizados. Os registros são resistentes a adulteração computacional com a existência de links criptográficos;
       
       \item Link criptográfico: o link criptográfico entre cada registro é classificado em ordem cronológica construindo uma cadeia de integridade pela blockchain. A assinatura digital verifica a integridade de cada registro usando técnicas de hashing e criptografia de chave assimétrica. Violar a integridade do registro do bloco ou da transação acaba tornando o registro e o bloco inválidos.
       
    \end{itemize}
  
    A segurança da blockchain parte dos avanços da criptografia e do design e implementação da blockchain (Bitcoin, Ethereum, etc.). Foram realizadas propostas de blockchain, com o passar do tempo, para melhorar a eficiência da cadeia criptográfica de blocos, por exemplo, incorporar árvores Merkle e colocar vários documentos em um bloco \cite{ZhangSecurityBlock}. A blockchain foi construída para garantir diversas características em relação a segurança, como consistência, resistência a adulteração, pseudonimato e resistência a ataques de gasto duplo e Distributed Denial of Service (DDoS). Porém, mesmo com o nível atual de segurança que a blockchain consegue prover, em alguns cenários ainda faltam propriedades adicionais de segurança e privacidade \cite{ZhangSecurityBlock}.
    
    
\section{Contratos Inteligentes}

	Os contratos inteligentes podem ser considerados como um programa que é executado quando condições predeterminadas são atendidas (autoexecutável) e que está implantado na blockchain, podendo ser utilizado em serviços financeiros, saúde e governo. É capaz de suportar funções e mecanismos programáveis complexos para automatizar acordos e outros tipos de fluxos \cite{SHI2020101966}.
	
	Esse tipo de contrato, que pode ser utilizado na blockchain, permite que as partes possam fazer o uso dele para criar terceiros virtuais de confiança que se comportaram de acordo com as regras acordadas entre ambos, dessa forma permitindo a criação de protocolos complexos com um risco de não cumprimento muito baixo \cite{Arbitrum}.
	
	Na rede blockchain do Ethereum os contratos inteligentes são desenvolvidos formalmente em código de alto nível através do Solidity (linguagem de programação) e são compilados para serem executados pela Máquina Virtual Ethereum (EVM). No conceito do Solidity os contratos inteligentes são um conjunto de códigos, dados, funções e estados, que estão em um específico endereço na rede blockchain do Ethereum \cite{solidity}. 
    
    Para que uma conta interaja com um contrato ou para que ocorra interações entre contratos, deve-se ter o nome e os argumentos da função, assim surge a Application Binary Interface (ABI) que é  uma lista das funções e argumentos do contrato organizado no formato de um JavaScript Object Notation (JSON) , assim que ele é compilado. Dessa forma, utiliza-se da ABI para fazer o hash da definição da função e então criar o bytecode EVM necessário para chamar a função  \cite{solidityABI}.
    
\section{Algoritmos de Criptografia} 
    
    Os algoritmos criptográficos se dividem em dois tipos, os de chave simétrica ou chave privada, que ainda podem ser divididos em algoritmos que operam em um único bit ou em grupos de bits, e os de chave assimétrica ou chave  pública. A criptografia de chave pública foi criada em 1976 quando W. Diffie e M. Hellman,  propuseram  esta  nova  ideia,  a qual foi  seguida  por R. L. Rivest, A. Shamir, e L. Adleman, os criadores do algoritmo RSA \cite{RSAECCanalise}. O surgimento da  criptografia  baseada  em  curvas  elípticas, que é  um  tipo especial  de chave  pública, foi  inicialmente proposta por Miller e Koblitz no final dos anos 1980, a qual foi baseada em algoritmos e aplicações de chave pública até então já existentes \cite{RSAECCanalise}.
    
    Os algoritmos de chave simétrica vão possuir uma chave (privada) que é igual para ambas as partes que estão trocando informações e deve permanecer em segredo, pois ela será utilizada para criptografar e descriptografar as informações, é um método simples que facilita a criptografia entretando o problema dessa forma reside no compartilhamento da chave entre as partes, pois se alguém conseguir interceptar essa troca e obter acesso a chave a pessoa terá acesso para criptografar e descriptograr as informações \cite{articleSime}. Por outro lado os algoritmos de chave assimétrica segundo \citeonline{ECCRSAperf}, são algoritmos que dependem do uso de uma chave pública e de uma chave privada. A chave pública será livremente distribuída sem comprometer de alguma maneira a chave privada, a qual deve ser mantida em segredo. A chave pública é utilizada para criptografar mensagens de texto simples e verificar assinaturas, já a chave privada é usada para assinar mensagens e descriptografar os textos criptografados para obter as mensagens em texto simples \cite{ECCRSAperf}. 
    
    Dessa forma, pode-se observar que a criptografia acaba sendo o mecanismo principal para se conseguir garantir a segurança dos dados nos sistemas de informação atuais. No trabaho será utilizado o Advanced Encryption Standard (AES) em conjunto com o RSA ou o ECC, para assim criar um criptossistema híbrido com objetivo de aumentar a complexidade e força da criptografia. A seguir será abordado os algoritmos RSA, ECC e AES.
	
\subsection{RSA}
    
    O algoritmo de criptografia RSA de chave assimétrica tornou-se o padrão para criptografia de chave pública sendo amplamente utilizado. Sua segurança reside no problema de fatoração de inteiros e o seu processo de descriptografia não é eficiente como seu processo de criptografia \cite{mahto2017rsa}. Para uma segurança de dados melhor e mais forte, o RSA acaba necessitando de  tamanhos de chave maiores, o que implica em mais sobrecarga sobre os sistemas. Dessa forma, para os sistemas que possuem restrição de memória, o RSA se torna uma segunda opção \cite{mahto2017rsa}.
    
    O funcionamento do algoritmo RSA se baseia em números primos, assim quanto maior for o número escolhido mais segura a chave será. O cálculo das chaves pode ser feita através dos seguintes passos segundo \citeonline{hackingnaweb}:
    
    \begin{itemize}    
        \item Escolher dois números primos: os números escolhidos em uma implementação real serão muito maiores como o RSA de 2048-4096 bits, porém com o intuito de facilitar os cálculos e o entendimento os numeros serão \textbf{P = 3} e \textbf{Q = 11};
       
        \item Calcular o produto dos dois números \textbf{P} e \textbf{Q}: calculando \textbf{N = 3 * 11 = 33}, esse resultado será utilizado depois na cifragem e na decifragem;
       
        \item Calcular a função totiente de Euler: a fórmula \textbf{totiente(N) = (P - 1) * (Q - 1)} é utilizada para calcular o número de coprimos à um número, sendo que coprimo é um número que o máximo divisor comum entre ele e outro número é 1. Calculando com o número do exemplo \textbf{totiente(33) = (3 – 1) * (11 – 1) = 2 * 10 = 20}, assim o \textbf{N} possui 20 coprimos;
        
        \item Escolher um número \textbf{E} que seja um dos coprimos de \textbf{N}: dos coprimos de \textbf{N} temos 1, 2, 4, 5, 7, 8, 10, 13, 14, 16, 17, 19, 20, 23, 25, 26, 28, 29, 31 e 32, assim pode-se escolher, por exemplo, o número 17 para ser o valor de \textbf{E} que vai compor a chave pública;
        
        \item O \textbf{E} e o \textbf{D} devem ser tais que a seguinte fórmula seja verdadeira: \textbf{E * D mod totiente(N) = 1}. Portanto assumimos que \textbf{17 * D mod 20} precisa ser igual a \textbf{1}. Realizando o calculo obtêm-se que o valor de \textbf{D = 13} satisfaz a condição do resto da divisão ser igual a um.
        
    \end{itemize}
    
    Finalizando esses passos, se obtêm tudo que é necessário para começar a criptografar através do algoritmo RSA:
    
    \begin{itemize}    
        \item A chave pública é: \textbf{E} = 17, \textbf{N} = 33;
       
        \item A chave privada é: \textbf{D} = 13, \textbf{N} = 33.
    \end{itemize}

    Para realizar a criptografia, por exemplo da mensagem ``9'', será utilizada a seguinte fórmula: \textbf{Crip = M$^E$ mod N}. No exemplo, \textbf{Crip = 9$^1$$^7$ mod 33 = 16677181699666569 mod 33 = 15}, assim o texto criptografado é ``15''.
    
    Para realizar a descriptografia do texto ``15'', que só é possivel com a chave privada, será utilizada a seguinte fórmula: \textbf{Decrip = Crip$^D$ mod N}. No exemplo, \textbf{Decrip = 15$^1$$^3$ mod 33 = 1946195068359375 mod 33 = 9}, assim o texto criptografado volta ao texto original ``9''.
    

\subsection{ECC}
    
    A segurança do algoritmo de criptografia ECC de chave assimétrica reside no uso das propriedades matemáticas da curva elíptica para realizar o cálculo das chaves criptográficas (problema do logaritmo discreto sobre curvas elípticas)\cite{mahto2017rsa}. É um sistema mais adequado e promissor para dispositivos que possuem restrição de memória (Smartphone e Smartcards), além de ser utilizado em blockchains, como por exemplo a do Bitcoin a blockchain com o maior valor de mercado atualmente. O ECC consegue manter os níveis de segurança de forma equivalente ao RSA e requer comparativamente menores parâmetros para criptografia e descriptografia do que o RSA \cite{mahto2017rsa}.
    
   Segundo \citeonline{mahto2017rsa}, uma curva elíptica \textbf{C} sobre um campo finito primo é definida por uma equação na forma \textbf{y$^2$ = x$^3$ + ax + b} satisfazendo a restrição \textbf{4a$^3$ + 27b$^2$ $\neq$ 0}. O cálculo das chaves pode ser feita através dos seguintes passos:
   
   \begin{itemize}    
        \item Definir os elementos públicos globais: \textbf{E\textsubscript{q} (a, b)} a curva elíptica com os parâmetros \textbf{a}, \textbf{b} e \textbf{q}, onde \textbf{q} é um primo ou inteiro na forma \textbf{2$^m$}. Selecione \textbf{G} que é o ponto na curva elíptica de ordem \textbf{n};
       
        \item Geração de chave do usuário A: selecione a chave privada \textbf{nA} sendo \textbf{nA < n}. Calcular a chave pública \textbf{PA} através da formula \textbf{PA = nA * G};
       
        \item Geração de chave do usuário B: selecione a chave privada \textbf{nB} sendo \textbf{nB < n}. Calcular a chave pública \textbf{PB} através da fórmula \textbf{PB = nB * G};
        
        \item Cálculo da chave secreta pelo usuário A: feita pela fórmula \textbf{K = nA * PB};
        
        \item Cálculo da chave secreta pelo usuário B: feita pela fórmula \textbf{K = nB * PA}.
        
    \end{itemize}
    
     Para realizar a criptografia pelo usuário A usando a chave pública do usuário B são realizados os seguintes passos:
     
    \begin{itemize}    
        \item Usuário A escolhe a mensagem \textbf{P\textsubscript{m}}, sendo \textbf{P\textsubscript{m}} um ponto (x, y) codificado com a ajuda da mensagem de texto simples \textbf{m}, ou seja, é o ponto usado para criptografar e descriptografar, e um número inteiro positivo aleatório \textbf{k};
       
        \item O texto cifrado será \textbf{C\textsubscript{m} = {k * G, Pm + k * PB} }.
        
    \end{itemize}
    
    Para realizar a descriptografia pelo usuário B usando a sua chave privada são realizados os seguintes passos:
     
    \begin{itemize}    
        \item O texto cifrado é \textbf{C\textsubscript{m}};
       
        \item O texto plano será \textbf{P\textsubscript{m} = P\textsubscript{m} + k * PB - nB * (k * G)}
        
        \hspace{4.15cm}\textbf{ = P\textsubscript{m}  + k (nB * G) - nB * (k * G)}.
        
    \end{itemize}

\subsection{AES}
    O algoritmo de criptografia AES de chave simétrica foi desenvolvido em 1998 por Joan Daemen e Vincent Rijmen, permite um tamanho de bloco de dados fixo de 128 bits e suporta tamanho de chave de 128, 192, e 256 bits, além de qualquer combinação de dados \cite{PATIL2016617}. Segundo \citeonline{articleSime} é um dos algoritmos de chave simétrica mais populares, sendo adotado como padrão pelo governo dos Estados Unidos, e considerado o substituto do Data Encryption Standard (DES), devido sua  rapidez, fácil execução e pouca exigência de memória. 
    
    Sendo a cifra de bloco de chave simétrica mais utilizada dentro da segurança de computadores, principalmente pela sua padronização pelo NIST e também por todas as cripto análises publicadas sobre este algoritmo, conseguindo resistir a diversos tipos de ataques \cite{saraiva2019prisec}. Assim, torna-se uma escolha ideal para a criptografia de dados de maior volume, devido a sua performance, podendo ser combinado com a segurança de um algoritmo de chave assimétrica.

\chapter{Tecnologias e Ferramentas}

\section{Rede Blockchain do Ethereum}
	
	Na blockchain do Ethereum, existe a EVM, que é um computador canônico, o qual todos os membros da rede Ethereum concordam. Cada nó da rede mantêm uma cópia do estado deste computador e qualquer participante pode transmitir uma solicitação de transação para que este computador execute \cite{ethereum}. Quando uma solicitação é realizada os outros participantes da rede verificam, validam e executam o cálculo, o qual causa uma mudança de estado da EVM que será confirmada e propagada por toda a rede ou pela a maioria da rede \cite{ethereum}.
    
    A rede do Ethereum é uma rede blockchain não-permissionada, ou seja, possui acesso  aberto  sem  a  necessidade  de autenticação ou a existência de uma entidade central, assim qualquer usuário pode criar uma carteira e fazer parte da rede, até podendo se tornar um nó minerador e tentar validar os blocos em troca de uma taxa paga em Ether (ETH) por esse processamento realizado \cite{wust2018you}.

    Os mecanismos criptográficos da rede vão garantir que as transações sejam verificadas como válidas e então adicionadas ao blockchain, não podendo ser adulteradas posteriormente devido a dificuldade de realizar alguma adulteração, além disso também garantem que todas as transações sejam assinadas e executadas com as devidas permissões \cite{ethereum}. O Ethereum utiliza o mecanismo de consenso POW, desse modo o usuário que deseja adicionar um novo bloco à cadeia deve resolver um problema computacional difícil que vai requerer muito poder de computação, esse processamento que é conhecido como mineração, uma tentativa e erro de força bruta,  quem calcular com sucesso o resultado será recompensado em ETH \cite{ethereum}.
    
    O ETH é a criptomoeda nativa do Ethereum, a qual possibilita um mercado para computação provêndo um incentivo econômico para que os participantes verifiquem e executem solicitações de transação e forneçam também recursos computacionais para a rede \cite{ethereum}. O participante que transmite uma solicitação de transação deve oferecer alguma quantidade de ETH para a rede como recompensa, que é conhecido como \textit{Gas Fees} (taxas de gás), o valor vai corresponder ao tempo necessário para fazer o cálculo, o que acaba evitando a ação de membros maliciosos da rede com intuito de atrapalharem a execução da blockchain com scripts que consomem muitos recursos \cite{ethereum}, pois eles terão que pagar por esse tempo de execução. A recompensa é concedida ao usuário (nó minerador neste caso) que eventualmente realizar o trabalho de verificar, executar, enviar e transmitir a transação para a rede blockchain.
    
    O Ethereum é a primeira plataforma de blockchain de código aberto que oferece uma linguagem de contrato inteligente completa para os desenvolvedores implantarem seus contratos e Decentralized Applications (DApp) ao custo de uma taxa paga à rede no momento da migração do contrato para a blockchain \cite{SHI2020101966}.

\section{IPFS} 

    O IPFS é um sistema de arquivos distribuído (peer-to-peer) para armazenar, compartilhar e acessar arquivos, sites, aplicativos e dados com objetivo de tornar a web atualizável, resiliente e mais aberta \cite{ipfs}.
    
    Existem três princípios fundamentais para compreender o IPFS segundo o site oficial do \citeonline{ipfs}:
    \begin{itemize}    
        \item Identificação única por meio de endereçamento de conteúdo: os endereços de acesso as informações são baseados no conteúdo e não na sua localização, assim quando é adicionado um novo arquivo no IPFS esse arquivo é dividido em pedaços chamados “IPFS’s objects” (protegidos por criptografia, sistema de hash e assinatura digital) que vão possuir um Content Identifier (CID), o qual é um registro permanente do arquivo, além disso os “IPFS’s objects” guardam até 256kb de dados e também podem conter links para outros objetos que estão na rede;
        
        \item Vinculação de conteúdo por meio de Directed Acyclic Graphs (DAG): a estruturada do IPFS é um grande DAG, especificamente um Merkle DAG onde cada nó tem um identificador que é o resultado do hash do conteúdo do nó. A representação de seu conteúdo é feita em blocos e cada bloco tem uma raiz Merkle, isso  significa que diferentes partes do arquivo podem vir de diferentes fontes. As Merkle DAGs só podem ser construídas a partir das folhas, cada nó em um Merkle DAG é a raiz de um (sub) Merkle DAG em si e os nós na Merkle DAG são imutáveis;
        
        \item Descoberta de conteúdo por meio de Distributed Hash Tables (DHT): o IPFS usa uma tabela de hash distribuída, que é uma tabela dividida entre todos os pares da rede descentralizada, para armazenar as chaves e valores. O projeto ``libp2p'' é a parte do IPFS que fornece o DHT e controla os pares que se conectam e conversam entre si. Utiliza-se o DHT para encontrar a localização atual dos pares que estão armazenando o conteúdo que está sendo buscado.

    \end{itemize}
    
    Por exemplo, quando outros nós vão procurar por um conteúdo eles perguntam para os nós pares quem está armazenando o conteúdo que possui um determinado CID, os nós começam a responder e assim é possível se ter acesso a localização do conteúdo que se deseja. Após encontrar essa localização o IPFS utiliza o Bitswap para solicitar e enviar blocos entre os pontos, permitindo realizar a conexão aos pares de peers que possuem o conteúdo que está sendo buscado. Os nós que baixarem um determinado arquivo da rede vão possuir uma cópia na cache, e dessa forma se tornam provedores daquele conteúdo, o qual podem fornecer por um tempo ou descartar para limpar espaço de memória \cite{ipfs}.
    
    Portanto o IPFS permite que os membros da rede armazenem apenas o conteúdo que estão interessados, além de algumas informações de indexação que são utilizadas para ajudar a descobrir qual nó está armazenando que conteúdo. Além disso, se for adicionado uma nova versão de um arquivo, será feito um novo hash diferente e com um novo CID, assim garantindo a resistência a adulteração e censura, pois não são sobrescritas as alterações no arquivo original, além da reutilização de partes comuns do arquivo para minimizar custos de armazenamento \cite{ipfs}. O IPFS utiliza o sistema de nomenclatura descentralizado InterPlanetary Name System (IPNS) e o DNSLink para mapear CIDs para nomes Domain Name System (DNS) legíveis por humanos e utiliza criptografia de transporte para proteger os dados enviados entre dois pontos  \cite{ipfs}.

    Para os desenvolvedores blockchain o endereçamento de conteúdo IPFS permite o armazenamento de grandes arquivos fora de uma blockchain específica, pois ao invés de armazenar o arquivo na blockchain, que pode ser custoso, pode-se armazenar apenas os links imutáveis e permanentes, que vão referenciar o arquivo armazenado dentro do IPFS, nas transações que serão persistidas na blockchain, possibilitando o carimbo de data e hora e a proteção do conteúdo sem ter que adicionar os dados de maior volume na própria blockchain \cite{ipfs}.

\section{Truffle Suite} 
    
    O Ganache, o Truffle e o Drizzle formam a coleção de ferramentas conhecida como Truffle Suite, que são softwares feitos especificamente para o desenvolvimento na blockchain que até grandes empresas como a Amazon e a Microsoft utilizam \cite{truffleSuite}. Dentre as três ferramentas serão abordados o Ganache e o Truffle, os quais serão utilizados para o desenvolvimento da DApp (Decentralized Application) proposta neste trabalho.
    
\subsection{Truffle}

    O Truffle é um ambiente de desenvolvimento, framework de teste e pipeline de ativos para blockchains usando a Ethereum Virtual Machine (EVM) \cite{truffle}.
    
    Segundo o site do \citeonline{truffle}, as funcionalidades que o desenvolvedor consegue usufruir ao utilizar o Truffle são:
    
    \begin{itemize}    
        \item Compilação, vinculação, implantação e gerenciamento binário integrado de contrato inteligente;
        
        \item Teste de contrato automatizado;
        
        \item Estrutura de implantação e migração programável e extensível;
        
        \item Gerenciamento de rede para implantação em redes públicas e privadas;
        
        \item Gerenciamento de pacotes com EthPM \& NPM, utilizando o padrão ERC190;
        
        \item Console interativo para comunicação direta do contrato;
        
        \item Pipeline de construção configurável;
        
        \item Executor de script externo que executa scripts em um ambiente Truffle.
        
    \end{itemize}

\subsection{Ganache}

	O Ganache permite o seu usuário ter uma blockchain pessoal para desenvolvimento de DApps possibilitando que o desenvolvedor desenvolva, implemente e teste seus DApps em um ambiente seguro e determinístico. É possível desenvolver para, por exemplo, a rede do Ethereum e Corda. Além disso, através do Ganache é possível testar como a DApp afeta a blockchain e examinar detalhes como suas contas, saldos, criações de contratos inteligentes e custos de gás \cite{ganache}.

\section{MetaMask} 
    
    O MetaMask é uma carteira criptografada (digital) e um gateway para aplicativos da blockchain, que possibilita aos usuários o gerenciamento de suas contas, chaves e tokens de várias maneiras, incluindo carteiras de hardware, além de isolar o usuário do contexto do site \cite{metamask}. Está disponível como uma extensão de navegador e como um aplicativo móvel. 
    
    Para os desenvolvedores, é possível interagir com a API do Ethereum (globalmente disponível) que identifica os usuários de navegadores compatíveis com web3 e sempre que acontecer uma solicitação de uma assinatura de transação, o MetaMask irá solicitar ao usuário uma confirmação dessa transação, bem como indicar o custo dela \cite{metamask}. O MetaMask já vem configurado com algumas conexões para a rede blockchain do Ethereum e para várias redes de teste através da API do Infura. Além disso, atualmente o MetaMAsk é compatível com qualquer outra blockchain (públicas e privadas) que exponha uma API JSON RPC (Remote Procedure Calling) compatível com Ethereum \cite{metamask}.

\section{React} 

    O React é uma biblioteca JavaScript para construir interfaces de usuário. Essa biblioteca é declarativa, o que faz com que seu código seja mais previsível e simples de depurar, baseada em componentes o que torna fácil passar diversos tipos de dados ao longo da sua aplicação e ainda assim manter o estado fora do Document Object Model (DOM) \cite{react}. Os componentes do React implementam um método “render()'' que vai receber os dados de entrada e retornar o que deve ser exibido, além disso um componente pode manter os dados do estado interno \cite{react}. O React facilita a interface com outras bibliotecas e frameworks.
     
    
    O React faz uso do JavaScript XML (JSX) que não é obrigatório, sendo uma extensão de sintaxe para JavaScript que produz “elementos” do React e que facilita o desenvolvimento de aplicações ao mostrar, por exemplo, mensagens mais úteis de erro e aviso. O uso do JSX está no motivo do React adotar o fato de que a lógica de renderização é inerentemente acoplada com outras lógicas de interface de usuário (manipulação de eventos, mudança de estado com o tempo, preparação de dados para exibição, etc.) \cite{reactjsx}. Os elementos React são objetos simples e o responsável por atualizar o DOM para exibir os elementos React é o React DOM, o qual compara um elemento novo e seus filhos com os anteriores e somente vai aplicar as modificações necessárias \cite{reactjsx}.

\section{Node.js}
    
    O Node.js é como um runtime JavaScript assíncrono e orientado a eventos, projetado para criar aplicativos de rede escaláveis. Os usuários do Node.js não precisam se preocupar com travamento de processos, pois quase nenhuma função no Node.js executa de forma direta a entrada e saída, assim o processo nunca é bloqueado, exceto quando for utilizado métodos síncronos da biblioteca padrão do Node.js para executar a entra e saída \cite{nodejs}. 
    
    Influenciado por sistemas como o Event Machine do Ruby e o Twisted do Python, o Node.js apresenta um loop de eventos como uma construção em tempo de execução, sendo que não existe uma chamada que vai iniciar o evento de loop, ele simplesmente entra no loop de eventos após executar o script de entrada e sai do loop de eventos quando não existem mais retornos de chamada a serem executados \cite{nodejs}.
    
    Mesmo que tenha sido projetado sem threads, é possível fazer o uso de vários núcleos de um ambiente, através dos processos filhos que podem ser gerados usando a API child\_process.fork(). Utilizando do módulo de cluster, que possibilita compartilhar soquetes entre processos e assim fazer o balanceamento de carga em seus núcleos \cite{nodejs}.
    
    Portanto, acaba sendo comum a escolha do Node.js para o desenvolvimento de aplicações escaláveis, e no caso do desenvolvimento de DApps essa plataforma também acaba sendo muito preferida em conjunto com o React, assim desenvolvendo o front-end e back-end em JavaScript.
    
\section{WEB3} 
    
    A WEB 3.0 possui como foco a descentralização, diferente da WEB 1.0 e WEB 2.0, além de trazer também algumas características adicionais como ser verificável, autogovernado, sem permissão e distribuído . Os aplicativos Web3 (DApps) são executados em redes descentralizadas, em blockchains ou até  mesmo uma combinação dos dois formando, por exemplo, um protocolo criptoeconômico, já que a criptomoeda desempenha um grande papel em muitos desses protocolos, já que fornece um incentivo financeiro (tokens) para os nós que quiserem participar da criação, governança, contribuição ou melhoramento de um projeto \cite{web3}.
    
    Os projetos que são desenvolvidos em cima desse sistema da Web3 acabam oferecendo uma variedade de serviços como computação, armazenamento, hospedagem entre outros serviços que eram fornecidos, principalmente, por provedores de nuvem. Os usuários que consomem esses serviços da Web3 pagam para usar o protocolo, mas nesse caso o dinheiro vai diretamente para os participantes da rede eliminando intermediários desnecessários \cite{web3}.

    No desenvolvimento da DApp do presente trabalho será utilizada a biblioteca Web3 que é empregada para conectar-se à rede Ethereum a partir de um aplicativo. A ABI é dada à biblioteca Web3, a qual usa para dar acesso programático ao contrato implantado, neste caso, na rede do Ethereum. Um objeto da classe Web3 deve ser instanciado para se utilizar de suas funções. Cada instância da biblioteca Web3 pode-se conectar a uma rede Ethereum diferente \cite{panda2021investigation}. A instância do Web3 vai requer uma camada de comunicação conhecida como provedor, que atua como um meio entre a biblioteca Web3 e a rede Ethereum, e cada provedor tem um conjunto de métodos para enviar ou receber uma solicitação da rede Ethereum\cite{panda2021investigation}.
 
\section{API do Infura} 
    
    A API do Infura é alimentada por uma arquitetura orientada a microsserviços que é escalonada dinamicamente fornecendo acesso instantâneo por Hypertext Transfer Protocol Secure (HTTPS) e WebSockets à rede Ethereum, ou seja, fornece uma infraestrutura para DApps de maneira fácil e rápida \cite{infura}. Através do Infura os desenvolvedores podem se conectar a Ethereum e IPFS via HTTPS e WebSocket com tempos de resposta e disponibilidade satisfatórios. Além disso, a plataforma disponibiliza um painel que mostra o desempenho do aplicativo e o uso da API, detalhando as solicitações de métodos específicos, tempo de uso e outras funcionalidades que podem ajudar o desenvolvedor durante a contrução da sua aplicação \cite{infura}.
    
    Algums recursos disponíveis pela API do Infura segundo o site da plataforma \citeonline{infura}:
    
    \begin{itemize}    
        \item Suporta mainnet e testnets via JSON-RPC compatível com o cliente, HTTPS e Windows Sharepoint Services (WSS);
        
        \item Funciona com as últimas atualizações de rede com garantia mínima de 99,9\% de tempo de atividade;
        
        \item Conecta seu aplicativo com uma linha de código sem sincronização e sem configurações complicadas;
        
        \item Permite configurar, monitorar e analisar seus aplicativos com o painel de controle do Infura;
        
        \item Acesso aos dados do nó do arquivo Ethereum disponíveis como um adicional;
        
        \item Acesso 24 horas por dia, 7 dias por semana a equipes de suporte especializado e a comunidade de desenvolvedores experientes do Infura.
    
    \end{itemize}

\chapter{DApp}\label{cap_exemplos}

\begin{figure}[h]
    	\centering
        \caption{Esquema da arquitetura}  \label{fig:ArquiteturaSimples}
        \includegraphics[width=0.77\textwidth]{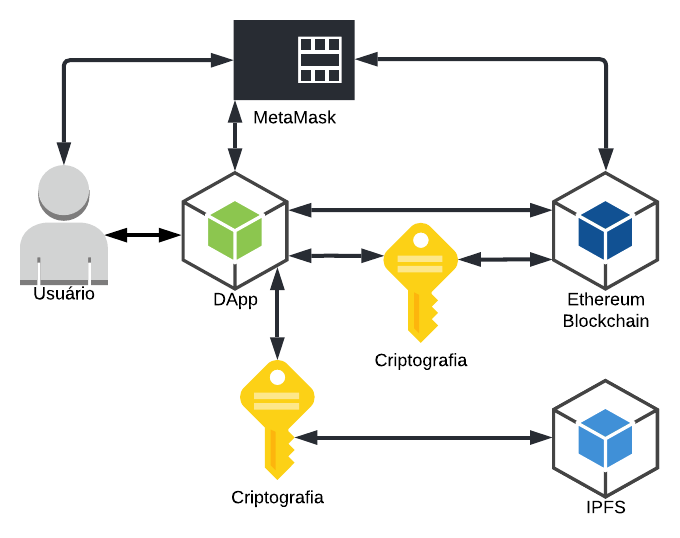}
        \legend{Fonte: Do autor, 2022}
    \end{figure}

   No projeto da DApp definiu-se uma arquitetura composta por seis componentes, conforme pode ser observado na  Figura \ref{fig:ArquiteturaSimples}. A partir desta arquitetura busca-se  o desenvolvimento de uma DApp que possa garantir a proteção da privacidade dos 
    dados de saúde armazenados em uma blockchain, usando mecanismos de criptografia.

 A seguir são descritos os componentes da arquitetura:
        
    \begin{itemize}    
    	\item{Usuário: vai realizar o acesso a aplicação, enviar e receber dados (informações, arquivos e permissões), além de confirmar as transações e pagar a taxa da persistência desses dados na blockchain}; 
    	\item{DApp: responsável por se comunicar com a interface da carteira digital (MetaMask), com a blockchain (Ethereum) enviando e recebendo os dados, com o banco descentralizado (IPFS)  enviando os arquivos e tratando o hash de retorno, com o usuário recebendo e mostrando os dados requisitados, e por fim,  realizar a criptografia/descriptografia do hash, da chave privada do AES, do arquivo e da chave privada compartilhada na concessão de permissão};
    	\item{Interface de Carteira Digital (MetaMask): responsável por gerenciar as transações realizadas na DApp, solicitando a confirmação do usuário e debitando a taxa para a persistência dos dados na blockchain};
    	\item{Blockchain (Ethereum): responsável por guardar os contratos inteligentes que regem como as informações pertinentes do arquivo e da permissão serão armazenadas, como por exemplo, o hash que é utilizado para se ter acesso ao arquivo no IPFS, o endereço do usuário que realizou uma transação e o endereço do usuário a qual está sendo cedida a permissão, realizando dessa forma o link do respectivo usuário a determinada informação dentro dos contratos inteligentes}; 
    	\item{Sistema de arquivos descentralizado (IPFS): responsável por receber os arquivos criptografados da DApp, armazená-los e devolver um respectivo hash único que indica onde se encontra essa determinada informação na rede do IPFS};
    	\item{Criptografia: esse é o componente chave para garantir a segurança e privacidade dos dados armazenados na arquitetura proposta, pois através dos algoritmos de criptografia RSA ou ECC a aplicação criptografa o hash e a chave privada do AES antes de enviar essas informações para a blockchain; Com o AES é feita a  criptografia do arquivo e da chave privada compartilhada na concessão da permissão. Dessa forma, somente o usuário que possui acesso a chave privada para descriptografar esse hash e a chave privada do AES, vai conseguir ter acesso às informações armazenadas. Este processo, garante que as informações na blockchain e no IPFS fiquem seguras}.
    \end{itemize}   

    Na sequência foram definidos os requisitos funcionais e não funcionais para a DApp.
    Os requisitos funcionais tem por finalidade descrever o comportamento da aplicação e as funções que devem ser previstas. Como requisitos funcionais, temos:   
    
    \begin{itemize}    
        \item{Realizar Login: a aplicação só estará disponível após login do usuário pelo MetaMask};
        \item{Informar chaves de criptografia: toda vez que o usuário logar na aplicação serão geradas suas chaves (pública e privada); a aplicação possui um espaço para que o usuário informe suas chaves; na primeira vez que acessar a aplicação pode guardar suas chaves da forma que preferir, e no próximo login pode informar as suas chaves};
        \item{Armazenar informações: o usuário pode enviar informações a serem armazenadas no IPFS, de forma segura através da criptografia AES, o qual gera um hash para se ter acesso ao conteúdo na rede, esse hash e a chave privada do AES, gerada aleatoriamente, serão armazenados de forma criptografada na blockchain do Ethereum, através de um dos algoritmos de chave assimétrica (ECC ou RSA), para garantir a privacidade e o controle do acesso aos dados}; 
        \item{Confirmar ou rejeitar transações: o usuário, após enviar uma requisição para armazenar informações ou uma permissão nos respectivos contratos, pode aceitar ou rejeitar a transação através do MetaMask};
        \item{Buscar informações: o usuário pode buscar as informações que possui armazenadas no IPFS através do hash armazenado na blockchain, de acordo com o seu usuário logado};
        \item{Conceder permissão: o usuário A pode conceder permissão de acesso a um arquivo que possui armazenado no IPFS para um usuário B, através de algumas informações como o hash armazenado na blockchain, o tempo específico que a permissão vai ter de validade, o endereço do usuário B e a chave pública do usuário B};
        \item{Visualizar arquivos com permissão: um usuário que teve uma permissão concedida por outro usuário a um arquivo, pode ter acesso para visualizar esse arquivo dentro do tempo que foi estipulado pelo usuário que cedeu a permissão}.
    \end{itemize}


	Já com relação aos requisitos não funcionais, que são premissas e/ou restrições técnicas para a aplicação, foram elencados:

   \begin{itemize}    
    	\item{DApp: a aplicação possuirá as características de uma aplicação descentralizada}; 
    	\item{Armazenamento da aplicação: a aplicação não armazena nenhum dado de forma permanente e centralizada, apenas temporariamente};
    	\item{Segurança dos dados: os arquivos enviados ao IPFS serão criptografados pelo algoritmo AES. O hash, que identifica o local dos dados, e  a chave privada do AES (utilizada para criptografar o arquivo)  serão criptografados utilizando o RSA ou o ECC, antes de serem armazenados na blockchain, assim permitindo que o usuário consiga acessar novamente os seus dados de forma segura. A chave privada compartilhada na permissão será criptografada pelo AES}.
    \end{itemize}

    Considerando a arquitetura da Figura \ref{fig:ArquiteturaSimples}, a seguir é detalhado o funcionamento para cada um dos casos de uso dos requisitos funcionais incorporados na Dapp:
    
		\begin{itemize}    
    	\item{1. Realizar Login: 
            \begin{itemize}    
            	\item{1. O usuário acessa a aplicação (DApp) que permite o envio/recebimento de arquivos, sendo necessário o usuário estar logado a uma interface Web3 de carteira digital no navegador, como o MetaMask}; 
            	\item{2. A DApp estabelece uma conexão com a interface de carteira digital (MetaMask) no navegador para realizar as transações com a blockchain};
            	\item{3. A interface conecta com a blockchain para o gerenciamento das transações}; 
            	\item{4. Assim é estabelecida a conexão da DApp com a blockchain para as transações};
            	\item{5. Por fim, é feita a conexão da DApp com a blockchain do Ethereum via Web3 para o envio e recebimento de dados}.
            \end{itemize}}
        \item{2. Informar chaves de criptografia: 
            \begin{itemize}    
            	\item{1. Quando o usuário realiza o login é gerado sua chave pública e privada}; 
            	\item{2. O usuário então pode pegar e salvar suas chaves da forma que achar melhor};
            	\item{3. Quando o usuário acessar novamente a DApp ele poderá fornecer suas chaves para a aplicação utilizar na criptografia e descriptografia das informações}. 
            \end{itemize}}
    	\item{3. Armazenar informações: 
            \begin{itemize}    
            	\item{1. O usuário logado na DApp envia os arquivos e informações a serem armazenados}; 
            	\item{2. A DApp criptografa com o AES os arquivos e envia ao IPFS};
            	\item{3. O IPFS retorna um hash que indica onde está o arquivo na sua rede};
            	\item{4. Antes da DApp enviar os dados, o hash do arquivo e a chave privada do AES são criptografados pela DApp utilizando RSA ou ECC};
            	\item{5. Após a criptografia, os dados são enviados para serem armazenados na blockchain, através do contrato de armazenamento}.
            \end{itemize}}
    	\item{4. Confirmar ou rejeitar transações: 
            \begin{itemize}    
            	\item{1. A blockchain solicita a confirmação da transação ao MetaMask}; 
            	\item{2. O MetaMask solicita para o usuário confirmar ou rejeitar a transação};
            	\item{3. Se a transação for confirmada é realizado o pagamento do ``gas'' (taxa) da rede, senão é cancelado o envio dos dados}; 
            	\item{4. Após a confirmação, as informações são adicionadas na blockchain}.
            \end{itemize}} 
    	\item{5. Buscar informações: 
            \begin{itemize}    
            	\item{1. Após o usuário logar na aplicação (como demonstrado na etapa de login), a DApp vai buscar suas informações armazenadas na blockchain}; 
            	\item{2. A blockchain retorna as informações armazenadas para a aplicação utilizar};
            	\item{3. O hash do arquivo e a chave privada do AES são descriptografados, possibilitando assim o acesso ao arquivo no IPFS};
            	\item{4. O usuário visualiza as informações e acessa o arquivo armazenado no IPFS através do hash}.
            \end{itemize}}
    	\item{6. Conceder permissão: 
            \begin{itemize}    
            	\item{1. Para o usuário A conceder uma permissão ele vai precisar informar uma descrição, o tipo do arquivo, o hash do arquivo, o tempo de validade da permissão, o endereço e a chave pública do usuário a qual será cedida a permissão, neste caso usuário B}; 
            	\item{2. Antes de fazer o envio da permissão, a aplicação vai buscar a chave privada do AES e o hash do arquivo que o usuário A está compartilhando, para assim descriptografar com a chave privada do usuário A e criptografar com a chave pública do usuário B fornecida pelo usuário A};
            	\item{3. Criptografada e protegida as informações é feito o envio dos dados para a blockchain, através do contrato de permissão}.
            \end{itemize}}
    	\item{7. Visualizar arquivos com permissão: 
            \begin{itemize}    
            	\item{1. O usuário B que recebeu uma permissão sobre um arquivo de outro usuário A vai poder visualizar o arquivo dentro do tempo estipulado pelo usuário A}; 
            	\item{2. Quando o tempo para visualizar o arquivo estiver válido, será feita a descriptografia da chave privada do AES e do hash com a chave privada do usuário B, assim permitindo com que o usuário B acesse e visualize o arquivo do usuário A armazenado no IPFS}.
            	\item{3. Terminado o tempo de validade da permissão o usuário B não vai ter mais acesso ao arquivo pela DApp}; 
            \end{itemize}}
    \end{itemize}

\begin{figure}
    \centering
    \caption{Diagrama de sequência para o login e busca de arquivos}  \label{fig:diagramaLogin}
    \includegraphics[width=1\textwidth]{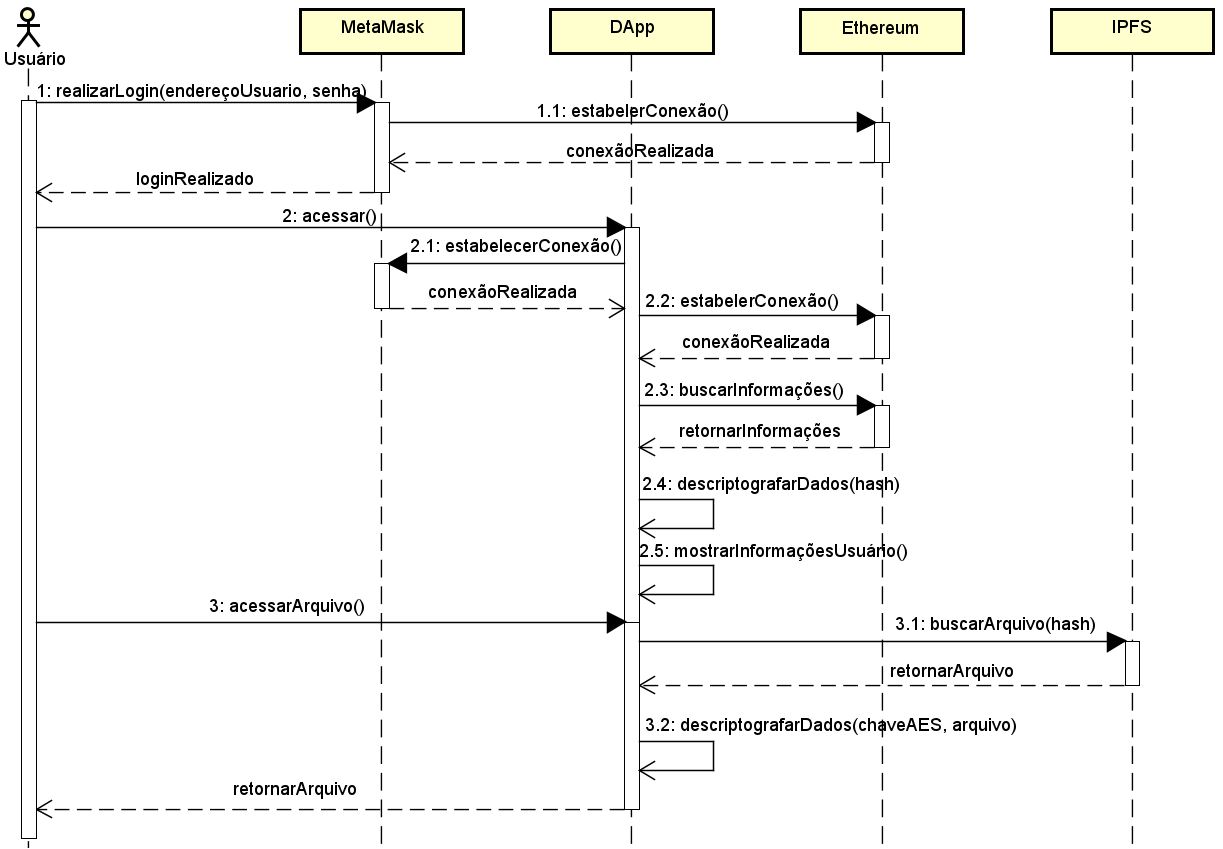}
    \legend{Fonte: Do autor, 2022}
    \end{figure}

    Para um entendimento de como os componentes do sistema interagem durante todo o processo, foram construídos quatro diagramas de sequência. O diagrama da \autoref{fig:diagramaLogin} apresenta as mensagens trocadas na interação dos objetos nos casos de usos realizar login e buscar informações. Inicialmente, é realizado o login no MetaMask para assim ser possível acessar a DApp. Acessando a DApp são estabelecidas as conexões necessárias, se o usuário possuir arquivos armazenados e permissões essas informações já são buscadas e mostradas para o usuário. Dessa forma, o usuário pode acessar um arquivo que será buscada através do hash no IPFS.

\begin{figure}
    \centering
    \caption{Diagrama de sequência para o envio de arquivos e gerenciamento das transações}  \label{fig:diagramaEnvio}
    \includegraphics[width=1\textwidth]{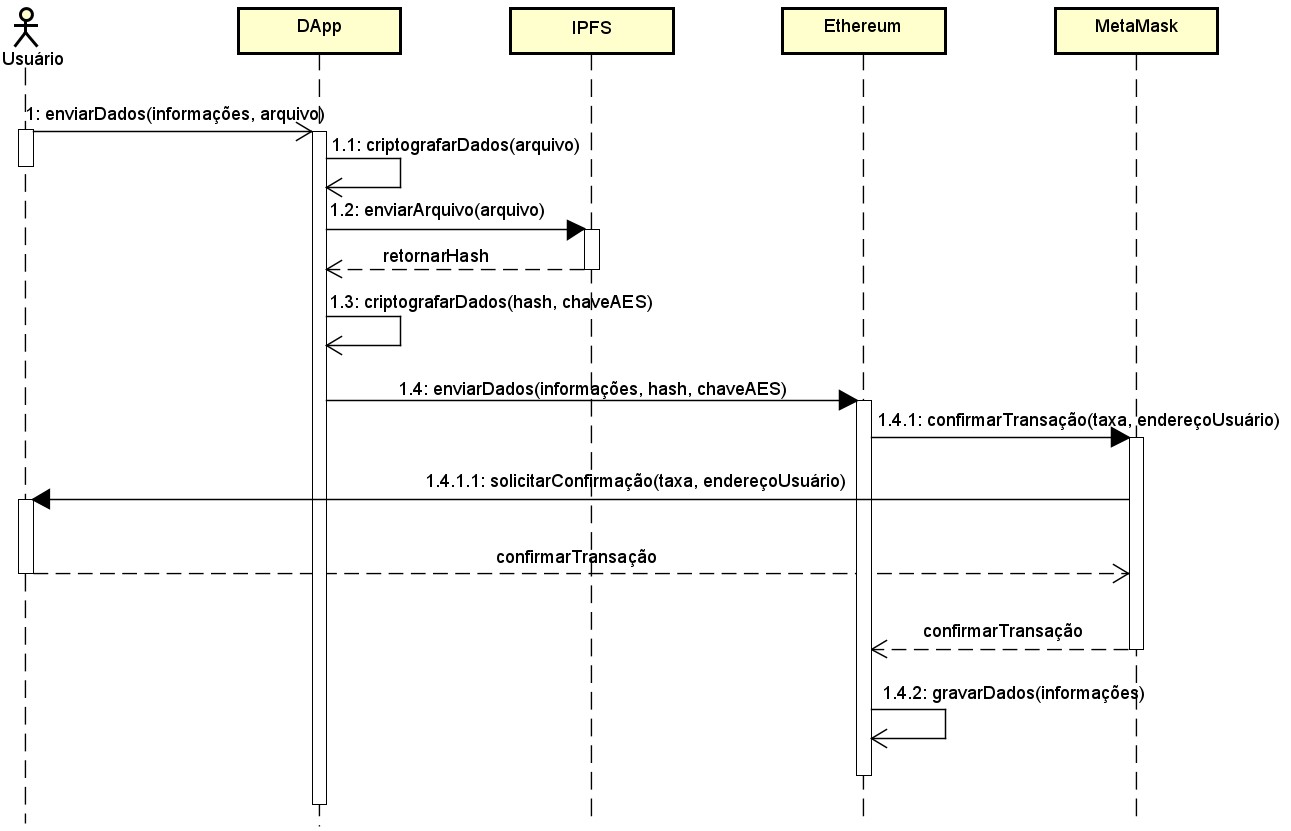}
    \legend{Fonte: Do autor, 2022}
    \end{figure}

    O diagrama da \autoref{fig:diagramaEnvio} traz a representação da sequência de
    operações dos casos de uso armazenar informações e confirmar ou rejeitar transações. Após o usuário ter realizado o login no MetaMask, pode realizar o envio de um arquivo a ser armazenado, podendo também enviar uma descrição. A DApp vai criptografar o arquivo através do AES e enviar para o IPFS, o qual vai retornar o hash que pode ser utilizado para encontrar esse conteúdo na rede do IPFS. A DApp faz a criptografia do hash e da chave do AES para assim enviar para a blockchain do Ethereum os dados. Quando é feita uma chamada para armazenar informações no Ethereum, o MetaMask vai confirmar com o usuário o pagamento da taxa para a rede, se for realizado o pagamento as informações são gravadas, senão o proceseso é cancelado.  
    
\begin{figure}
    \centering
    \caption{Diagrama de sequência para o envio de uma permissão}  \label{fig:diagramaPerm}
    \includegraphics[width=1\textwidth]{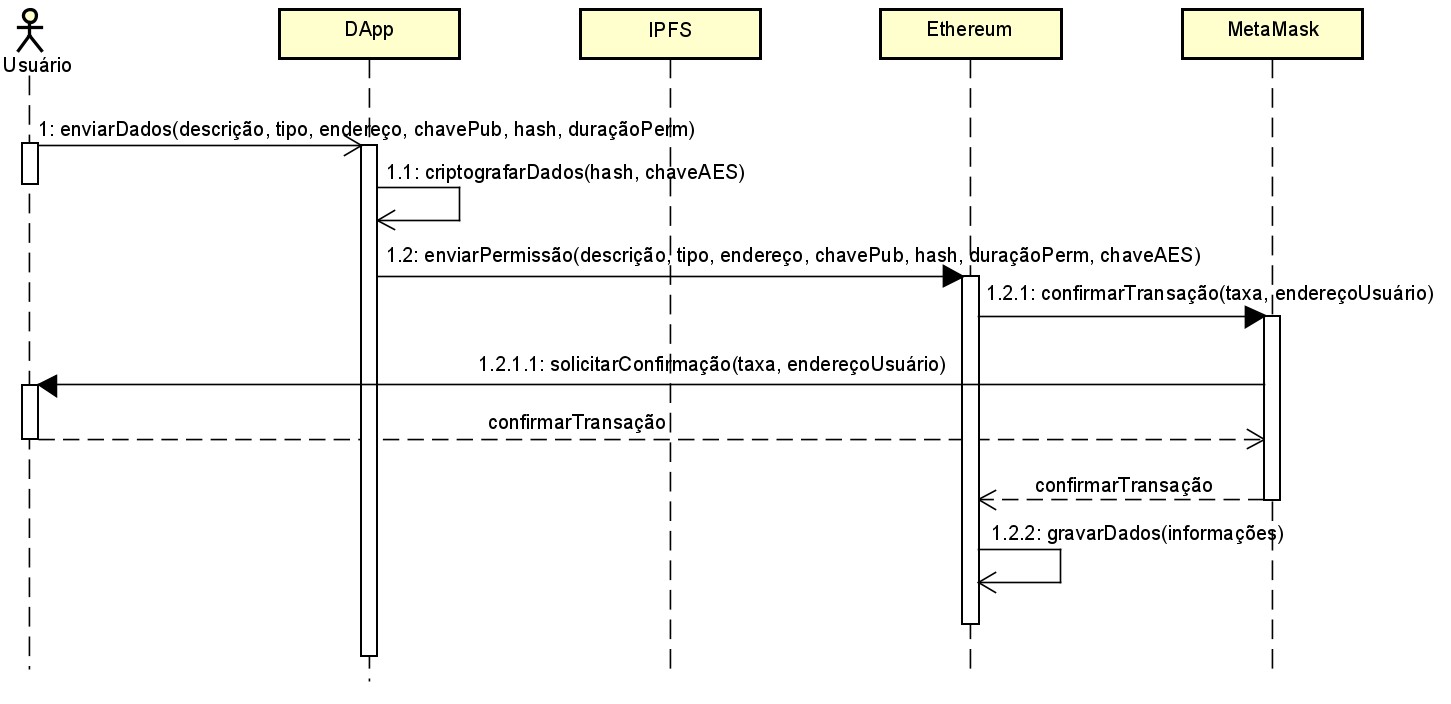}
    \legend{Fonte: Do autor, 2022}
    \end{figure}

    O diagrama da \autoref{fig:diagramaPerm} traz a representação da sequência de operações do caso de uso conceder permissão. Um usuário A pode enviar uma permissão de algum arquivo que tenha armazenado para outro usuário B, sendo necessário ser enviado essencialmente o endereço e a chave publica do usuário B, o hash do arquivo a ser compartilhado e o tempo de duração da permissão. Enviadas essas informações, a DApp vai criptografar o hash e a chave do AES do arquivo compartilhado com a chave pública do usuário B. Após a criptografia é feito o envio para a blockchain do Ethereum, também é necessário aceitar ou rejeitar a transação pelo MetaMask.
    
\begin{figure}
    \centering
    \caption{Diagrama de sequência para visualizar arquivos com permissão}  \label{fig:diagramaArqPerm}
    \includegraphics[width=1\textwidth]{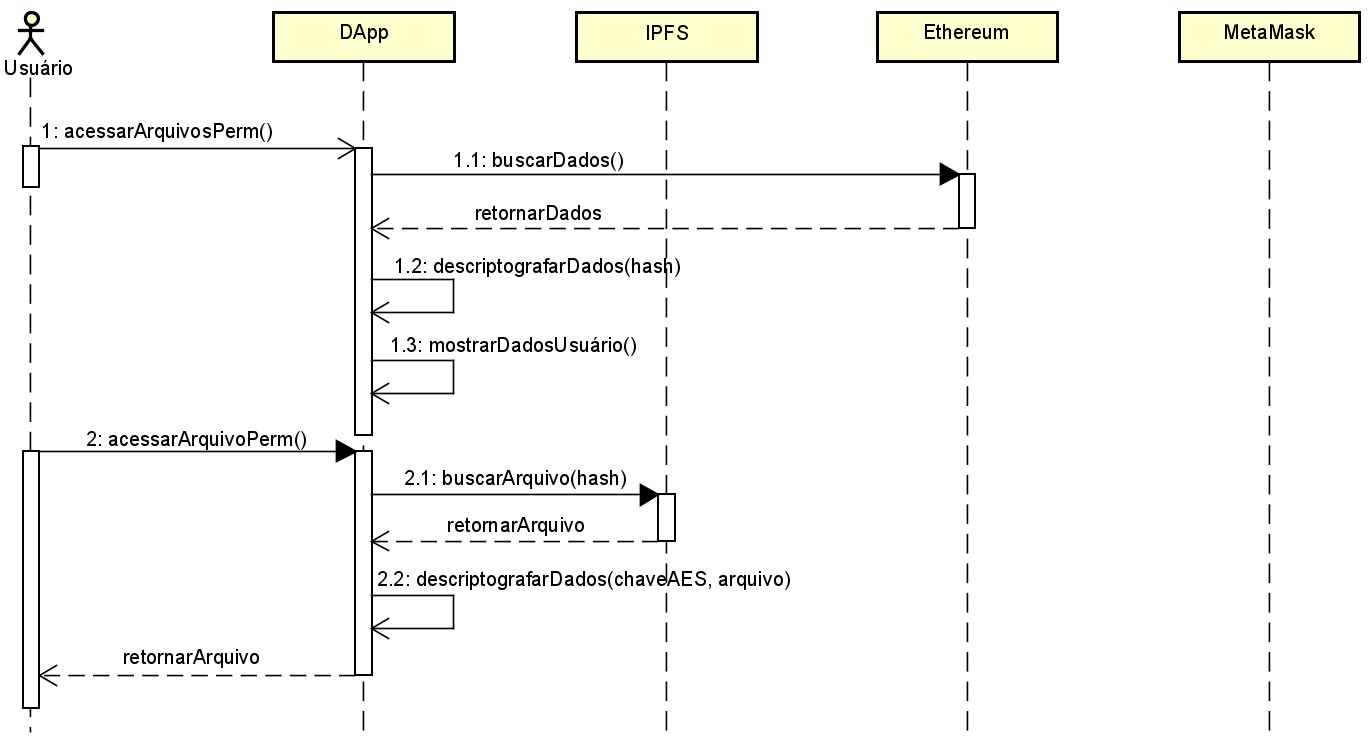}
    \legend{Fonte: Do autor, 2022}
    \end{figure}

    O diagrama da \autoref{fig:diagramaArqPerm} traz a representação da sequência de
    operações do caso de uso visualizar arquivos com permissão. Um usuário B que recebeu permissões sobre arquivos de outros usuários, vai conseguir acessar e visualizar dentro do tempo de duração da permissão concedida aos arquivos. Quando o usuário for acessar um arquivo, a DApp vai realizar a descriptografia do hash para buscar o arquivo no IPFS e da chave do AES para descriptografar o arquivo e permitir que o usuário B consiga visualizar.
    
\chapter{Implementações}\label{impl}
    
    Este capítulo apresenta detalhes quanto a implementação da DApp.     O front-end e o back-end da DApp foram desenvolvidos com JavaScript, HTML5, React e o Node.js, e a parte do contrato inteligente em Solidity. O código\footnote{Código disponível em: https://github.com/SegaBR/DApp} para a construção da DApp foi adaptado a partir de revisões feitas nas documentações das tecnologias descritas, alguns projetos de código aberto descentralizados desenvolvidos por \citeonline{dappUni} e também pesquisas nas documentações de bibliotecas de criptografia.
    
\section{Comunicação com a Blockchain}
    Inicialmente é necessário estabelecer a comunicação com um gateway para a rede blockchain do Ethereum, neste caso o Meta Mask, para assim estabelecermos a comunicação com a blockchain. Na \autoref{fig:ArquiteturaSimples} podemos visualizar a parte do código responsável por essa comunicação, sendo necessário importar para o projeto a biblioteca Web3.

\begin{figure}[h]
	\centering
    \caption{Comunicação com o MetaMask}  \label{fig:Web3Metamask}
    \includegraphics[width=1\textwidth]{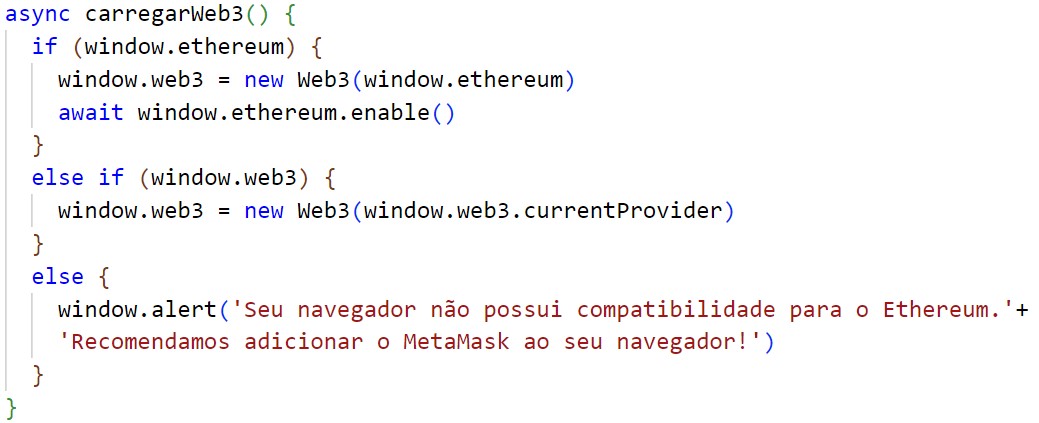}
    \legend{Fonte: Do autor, 2022}
\end{figure}

\begin{figure}[h]
	\centering
    \caption{Carregando a conta, buscando o ID da rede e dados do contrato}  \label{fig:ethereumDados}
    \includegraphics[width=0.8\textwidth]{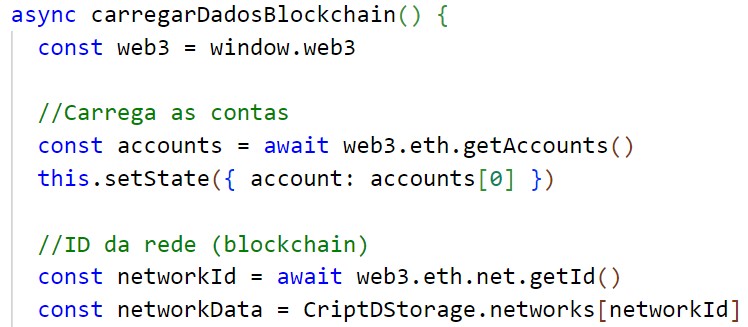}
    \legend{Fonte: Do autor, 2022}
\end{figure}

    Depois de estabelecer a comunicação anterior, é realizado o carregamento dos dados da blockchain. Na \autoref{fig:ethereumDados} é feita a busca pelas contas conectadas, pelo identificador da rede blockchain e pelos contratos implantados na rede, que neste caso são o CriptDStorage, que é o contrato responsável pela lógica de armazenamento, e o CriptDPermission, que é o contrato responsável pela a lógica das permissões.
    
\begin{figure}[h]
	\centering
    \caption{Carregando dados do contrato}  \label{fig:contratoDados}
    \includegraphics[width=1\textwidth]{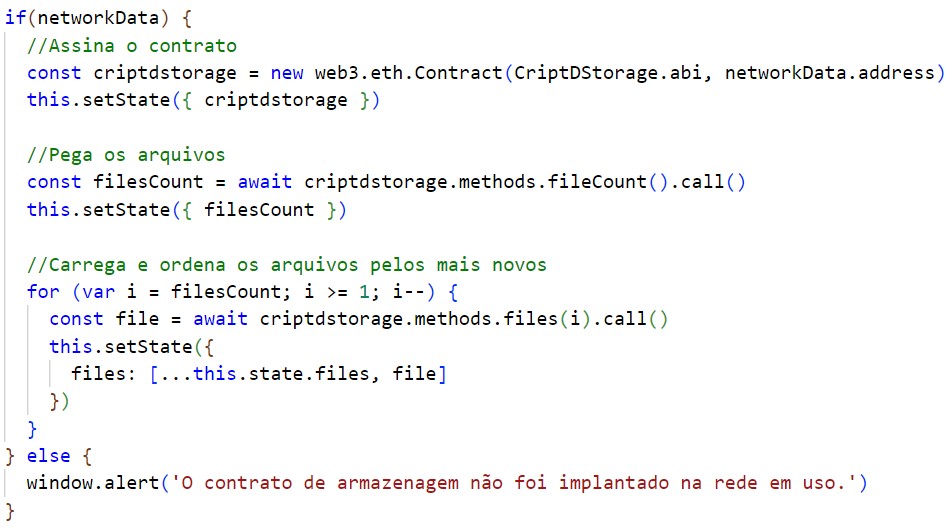}
    \legend{Fonte: Do autor, 2022}
\end{figure}

    Na \autoref{fig:contratoDados} é realizada a busca pelos dados armazenados e relacionados com cada contrato, no caso deste código,  está sendo feita a busca dos dados do contrato CriptDStorage responsável pela lógica de armazenagem do arquivo.

\begin{figure}[h]
	\centering
    \caption{Enviando dados ao contrato}  \label{fig:chamada_contrato1}
    \includegraphics[width=1\textwidth]{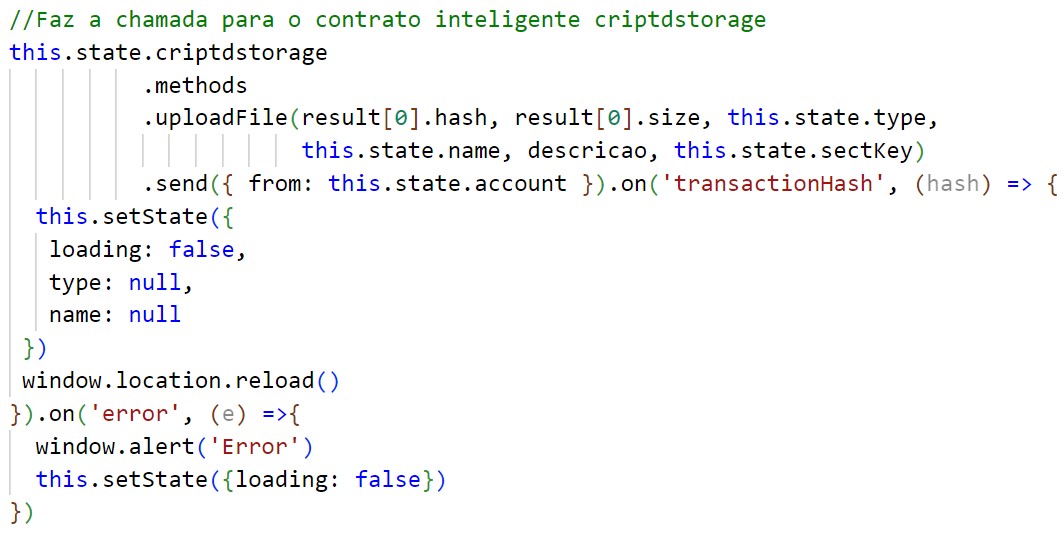}
    \legend{Fonte: Do autor, 2022}
\end{figure}

    Na \autoref{fig:chamada_contrato1} é realizada uma chamada ao contrato inteligente CriptDStorage, que faz um envio de armazenamento de um arquivo para a blockchain, chamando a função "uploadFile" do contrato inteligente e enviando as informações necessárias que a função vai utilizar. O método "send" que é chamado vai indicar qual é o usuário que está fazendo a requisição e que vai realizar o pagamento da taxa necessária para a rede do Ethereum persistir os dados. A transação é finalizada quando é gerado o hash da transação.

\section{Comunicação com o IPFS}
\begin{figure}[h]
	\centering
    \caption{Inicialização do IPFS}  \label{fig:iniciaIPFS}
    \includegraphics[width=1\textwidth]{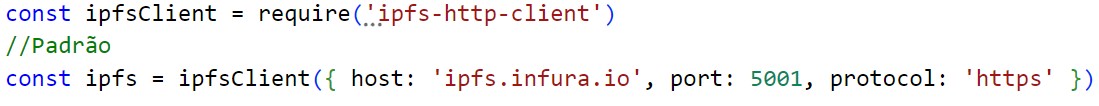}
    \legend{Fonte: Do autor, 2022}
\end{figure}

    Para estabelceler a comunicação com o IPFS primeiramente é feita uma inicialização, como pode ser observado na \autoref{fig:iniciaIPFS}, onde são informados os dados para estabelecer a comunicação que será realizada através da API do Infura.

\begin{figure}[h]
	\centering
    \caption{Envio de arquivo para o IPFS}  \label{fig:enviaIPFS}
    \includegraphics[width=0.7\textwidth]{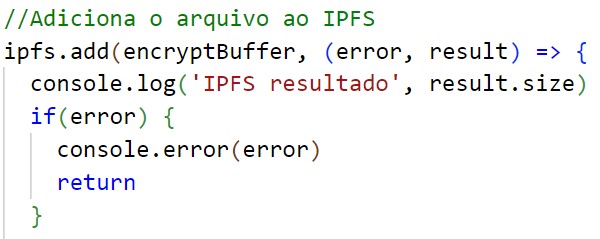}
    \legend{Fonte: Do autor, 2022}
\end{figure}

    Através do código da \autoref{fig:enviaIPFS} podemos realizar o envio de um arquivo para ser armazenado no IPFS, a variável \emph{result} retorna o hash, o qual é o CID que é utilizado para buscar esse conteúdo dentro da rede do IPFS.  
    
\begin{figure}[h]
	\centering
    \caption{Busca arquivo no IPFS}  \label{fig:buscaIPFS}
    \includegraphics[width=0.7\textwidth]{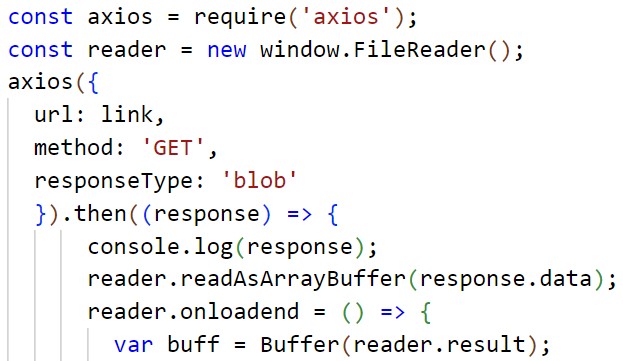}
    \legend{Fonte: Do autor, 2022}
\end{figure}

    Através do código da \autoref{fig:buscaIPFS} podemos realizar a busca de um arquivo armazenado no IPFS, através do link formado pelo hash, é feita a busca utilizando do \emph{Axios} que é um cliente HTTP baseado em promessas para o node.js que utiliza XMLHttpRequests. Assim, é recuperado o arquivo e transformado em um buffer para ser possível descriptografar e disponibilizar para o usuário acessar.  
    
\section{Criptografia}
    Em relação a criptografia implementada na aplicação ela é dividida dentre os três algortimos utilizados RSA, ECC e AES. 
    
\subsection{RSA}
\begin{figure}[h]
	\centering
    \caption{Gerar Chaves RSA}  \label{fig:gerarRSA}
    \includegraphics[width=0.85\textwidth]{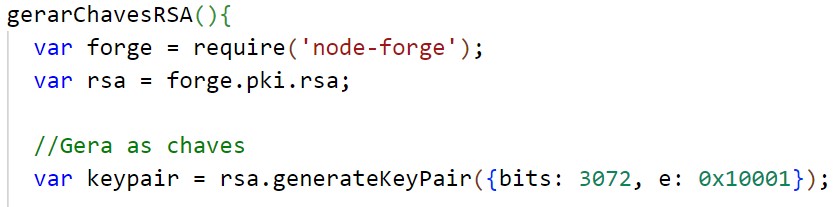}
    \legend{Fonte: Do autor, 2022}
\end{figure}

    A \autoref{fig:gerarRSA} apresenta como é feita a geração das chaves de 3072 bits do algortimo RSA através da biblioteca \emph{node-forge}.  

\begin{figure}[h]
	\centering
    \caption{Criptografar RSA}  \label{fig:criptografarRSA}
    \includegraphics[width=0.8\textwidth]{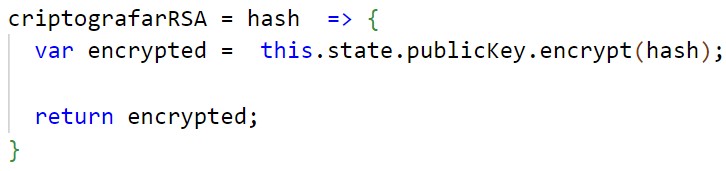}
    \legend{Fonte: Do autor, 2022}
\end{figure}

    O código da \autoref{fig:criptografarRSA} mostra a função que realiza a criptografia do dados com o RSA.
    
\begin{figure}[h]
	\centering
    \caption{Descriptografar RSA}  \label{fig:descriptografarRSA}
    \includegraphics[width=0.8\textwidth]{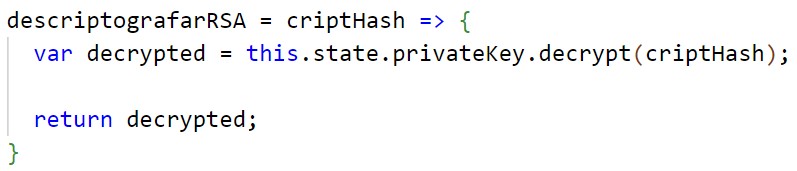}
    \legend{Fonte: Do autor, 2022}
\end{figure}

    Na \autoref{fig:descriptografarRSA} é mostrada a função que realiza a descriptografia com o RSA. As chaves ficam armazenadas temporariamente no state da aplicação.

\subsection{ECC}
\begin{figure}
	\centering
    \caption{Gerar Chaves ECC}  \label{fig:gerarECC}
    \includegraphics[width=0.7\textwidth]{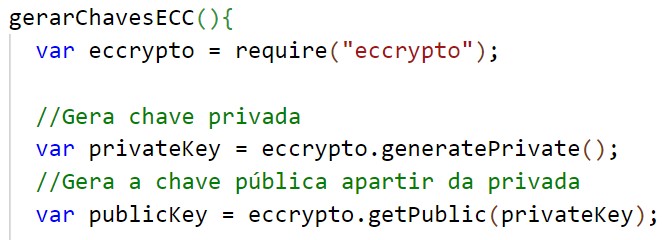}
    \legend{Fonte: Do autor, 2022}
\end{figure}

\begin{figure}
	\centering
    \caption{Criptografar ECC}  \label{fig:criptografarECC}
    \includegraphics[width=1\textwidth]{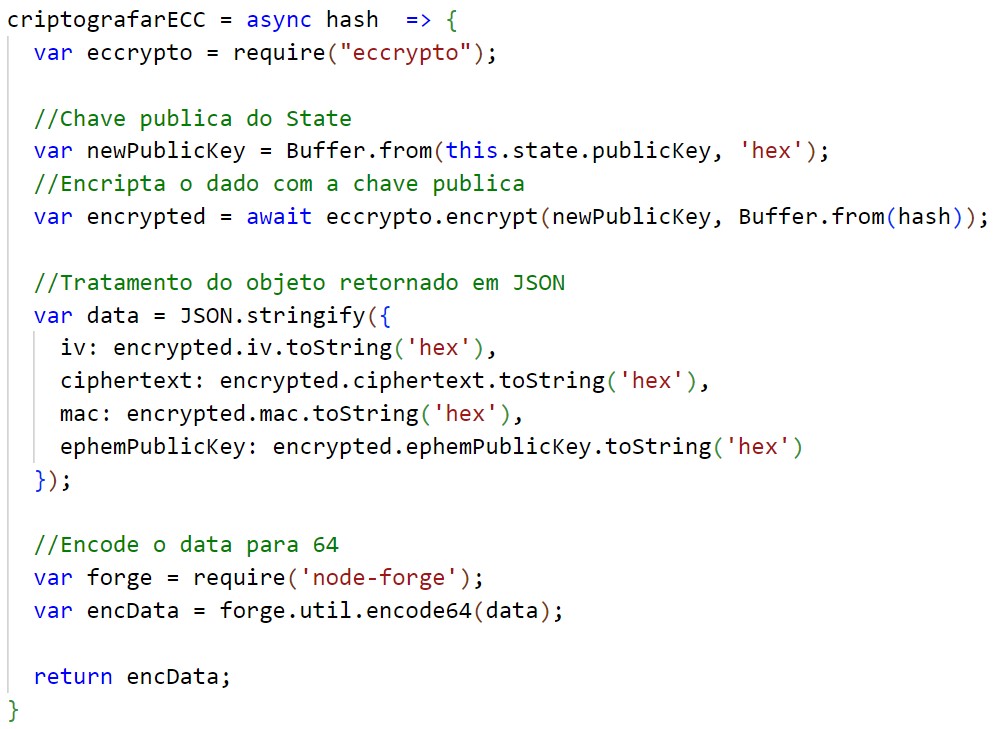}
    \legend{Fonte: Do autor, 2022}
\end{figure}

\begin{figure}
	\centering
    \caption{Descriptografar ECC}  \label{fig:descriptografarECC}
    \includegraphics[width=1\textwidth]{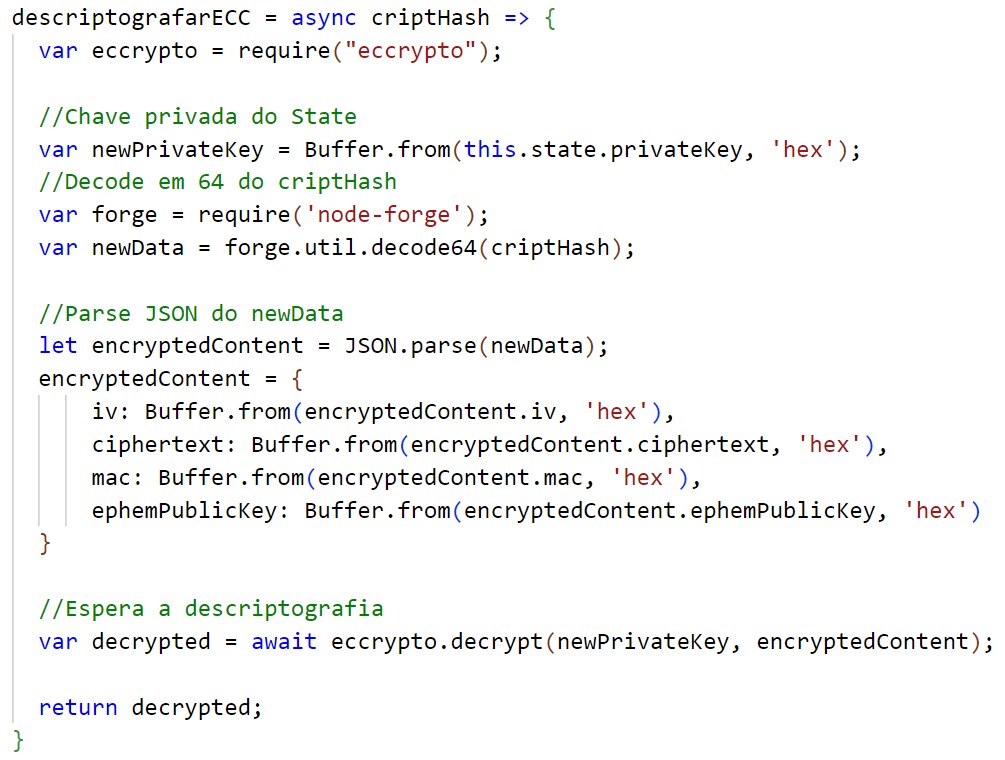}
    \legend{Fonte: Do autor, 2022}
\end{figure}

    O código apresentado na \autoref{fig:gerarECC} mostra a geração das chaves de 256 bits do algoritmo ECC através da biblioteca \emph{ecccrypto} utilizando a implementação do Elliptic Curve Integrated Encryption Scheme (ECIES) e da curva elíptica secp256k12. Na \autoref{fig:criptografarECC} é mostrada a função que realiza a criptografia com o ECC. A \autoref{fig:descriptografarECC} possui o a função que realiza a descriptografia com o RSA. As chaves ficam armazenadas temporariamente no state da aplicação, devido como é implementado o ECC na biblioteca, é necessário fazer um tratamento do objeto retornado, neste caso em formato JSON e realizando um encode64 do resultado para compactar as informações.

\subsection{AES}
\begin{figure}
	\centering
    \caption{Criptografar AES}  \label{fig:criptografarAES}
    \includegraphics[width=1\textwidth]{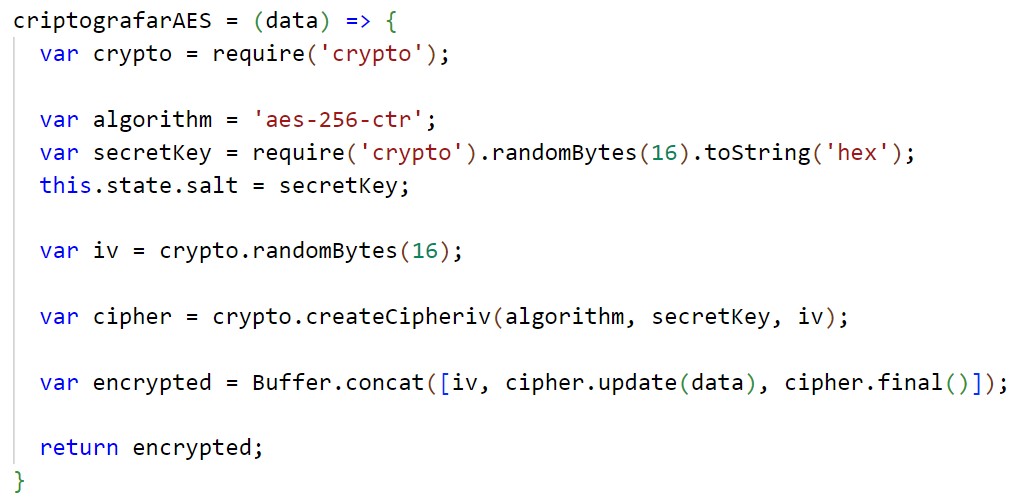}
    \legend{Fonte: Do autor, 2022}
\end{figure}

\begin{figure}
	\centering
    \caption{Descriptografar AES}  \label{fig:descriptografarAES}
    \includegraphics[width=1\textwidth]{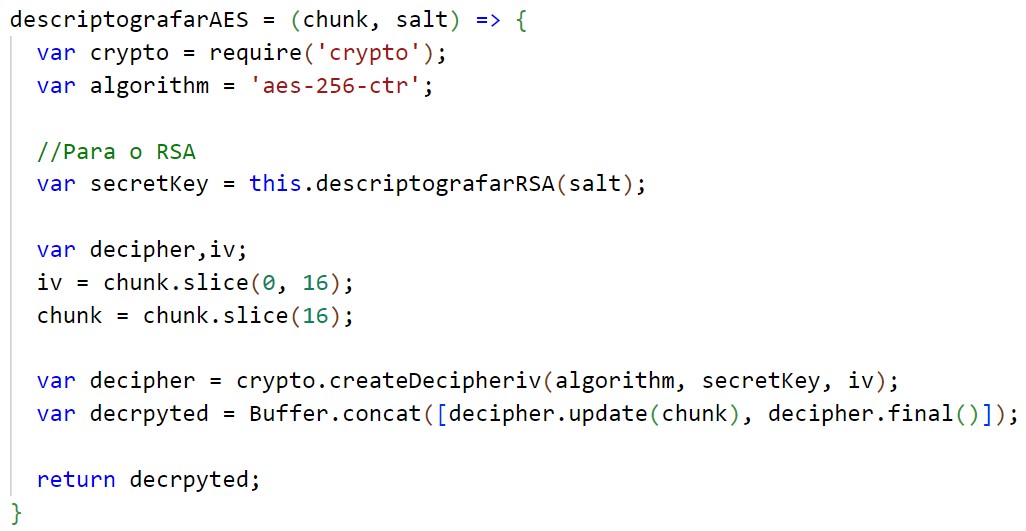}
    \legend{Fonte: Do autor, 2022}
\end{figure}

    Na \autoref{fig:criptografarAES} é mostrada como é realizado a criptografia do AES com tamanho de chave de 256 bits e através da biblioteca \emph{crypto}. A chave utilizada para criptografar os arquivos é gerada aleatoriamente também através da biblioteca \emph{crypto}. Na \autoref{fig:descriptografarAES} é realizada a descriptografia do AES e de acordo com o algoritmo utilizado, é feita a descriptografia da chave secreta.
    
\chapter{Interface da DApp}

\begin{figure}[h]
	\centering
    \caption{Tela do MetaMask}  \label{fig:telaMetamask}
    \includegraphics[width=0.4\textwidth]{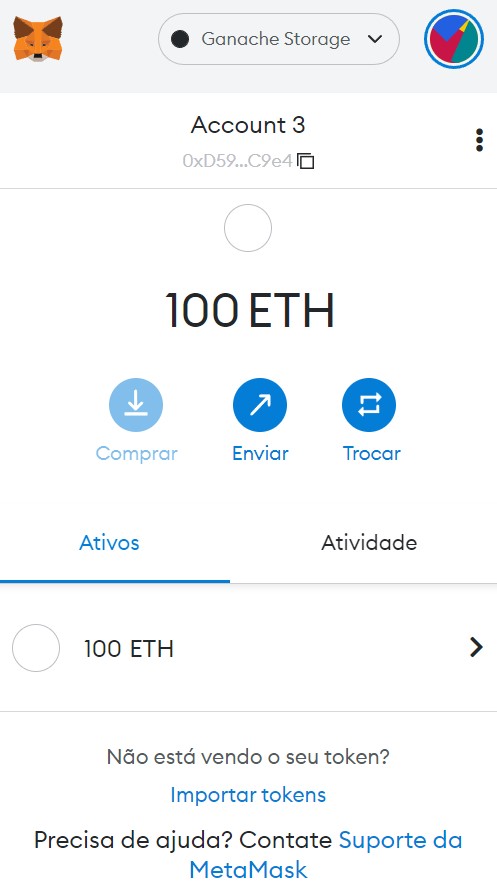}
    \legend{Fonte: Do autor, 2022}
\end{figure}

    Na \autoref{fig:telaMetamask} pode-se obervar a tela do MetaMask que é utilizada para geranciar as transações, já conectado na rede blockchain local simulada pelo Ganache. A interface do MetaMask trás informações de saldo da conta dos seus ativos, o seu endereço na rede e o histórico de transações, bem como outras configurações mais avançadas que podem ser feitas.
    
\begin{figure}[h]
	\centering
    \caption{Tela das chaves}  \label{fig:telaChave}
    \includegraphics[width=0.9\textwidth]{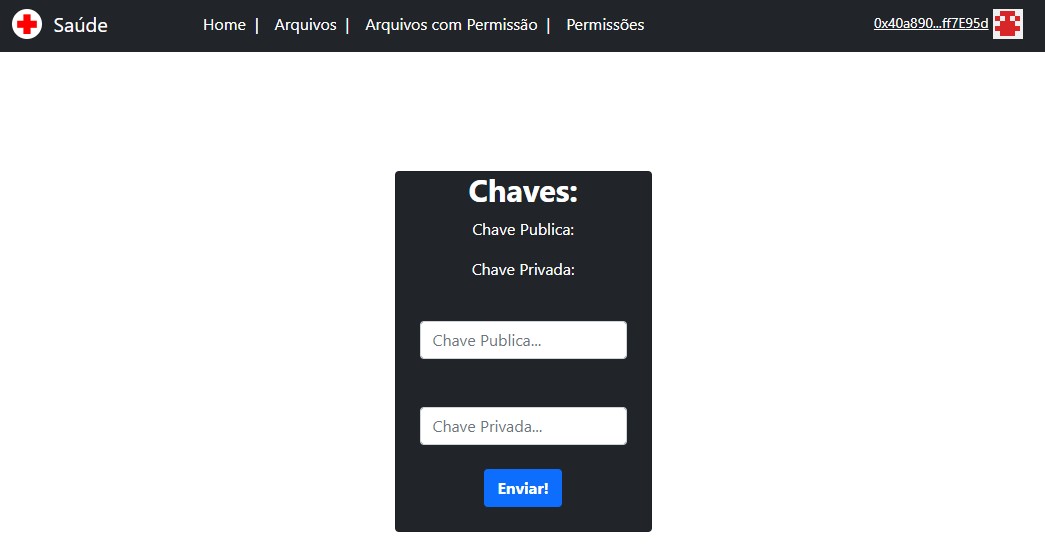}
    \legend{Fonte: Do autor, 2022}
\end{figure}

    A \autoref{fig:telaChave} apresenta a interface da DApp responsável por mostrar a chave pública e privada geradas para o usuário, o qual pode fazer uso delas como também informar e salvar outras chaves que ele está utilizando.
    
\begin{figure}[h]
	\centering
    \caption{Tela dos arquivos}  \label{fig:telaArquivo}
    \includegraphics[width=1\textwidth]{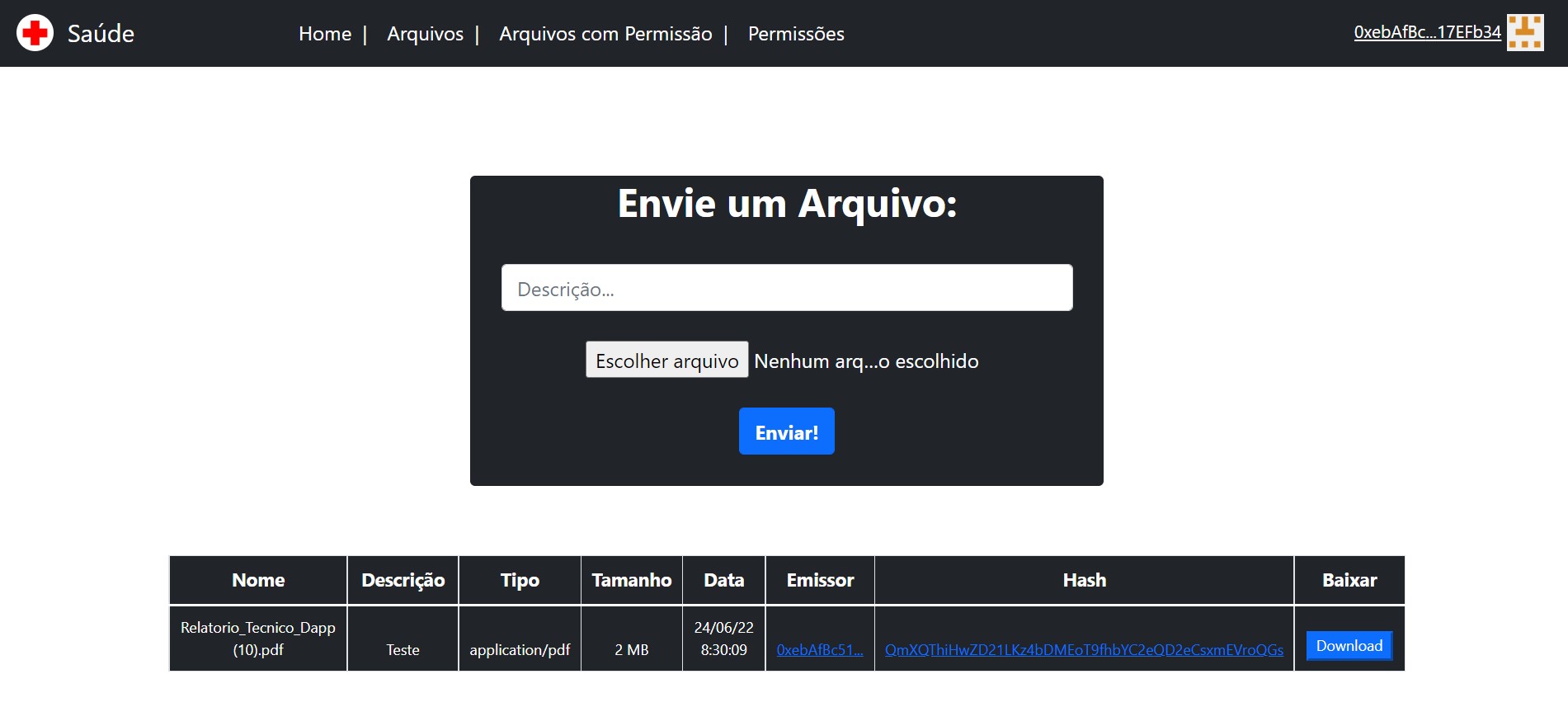}
    \legend{Fonte: Do autor, 2022}
\end{figure}

    Na \autoref{fig:telaArquivo} pode-se obervar a tela responsável por mostrar os arquivos que o usuário possui armazenados, sendo possível realizar o download, e também é onde ele pode realizar o envio de mais arquivos, sendo informada uma descrição e selecionado o arquivo que se deseja enviar.
    
\begin{figure}[h]
	\centering
    \caption{Tela das permissões}  \label{fig:telaPermissao}
    \includegraphics[width=0.8\textwidth]{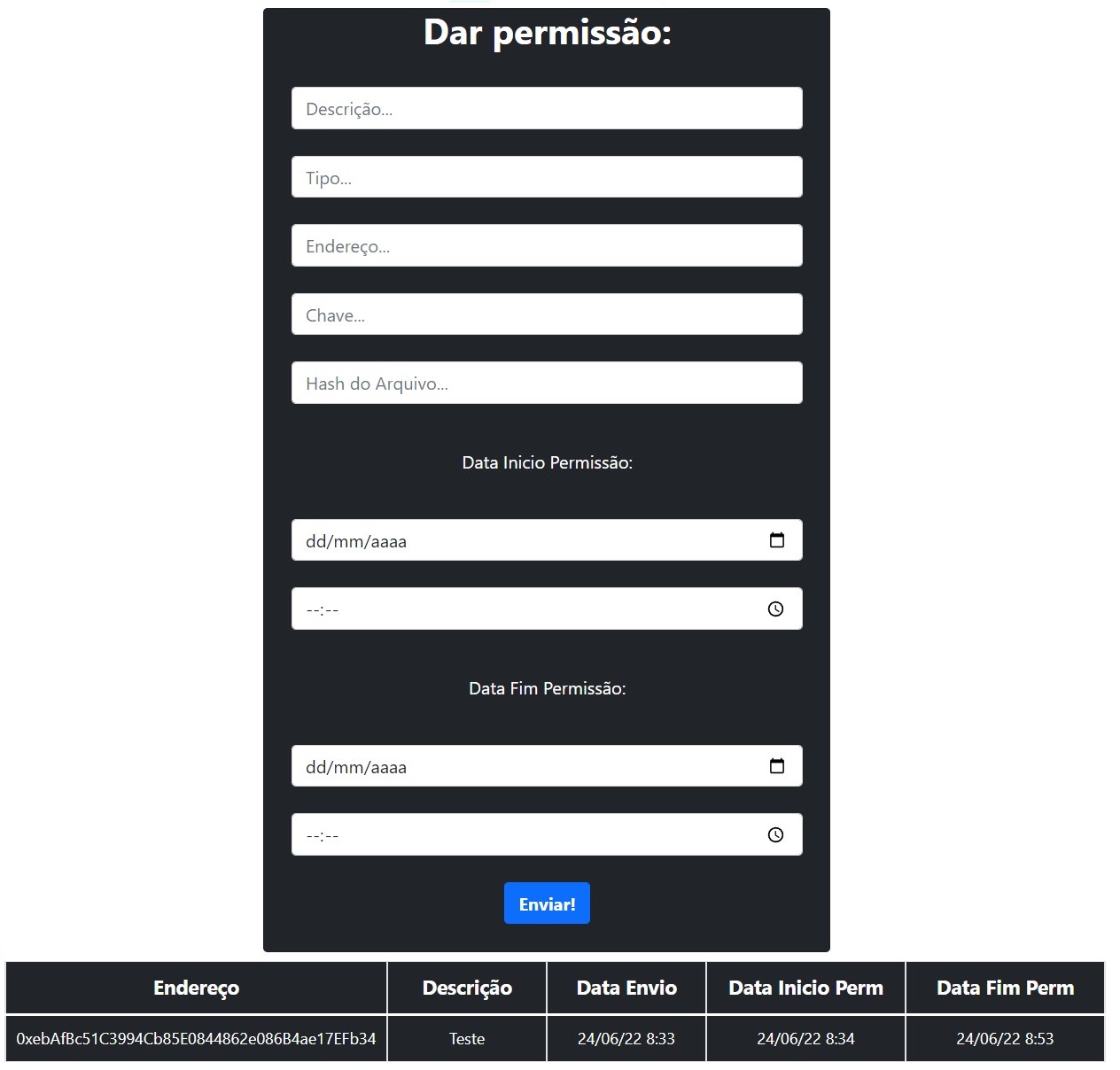}
    \legend{Fonte: Do autor, 2022}
\end{figure}

A interface responsável por mostrar as permissões que o usuário já enviou e para realizar o envio de mais permissões está na \autoref{fig:telaPermissao}. O usuário pode informar uma descrição, o tipo do arquivo, o endereço do usuário B que vai receber a permissão, a chave pública do usuário B, o hash do arquivo que está sendo compartilhado e, por fim, a data do início que a permissão vai começar a ser válida e a data do fim da válidade da permissão, ou seja, o tempo de duração da permissão.
    
\begin{figure}[h]
	\centering
    \caption{Tela dos arquivos com permissões}  \label{fig:telaPermArq}
    \includegraphics[width=1\textwidth]{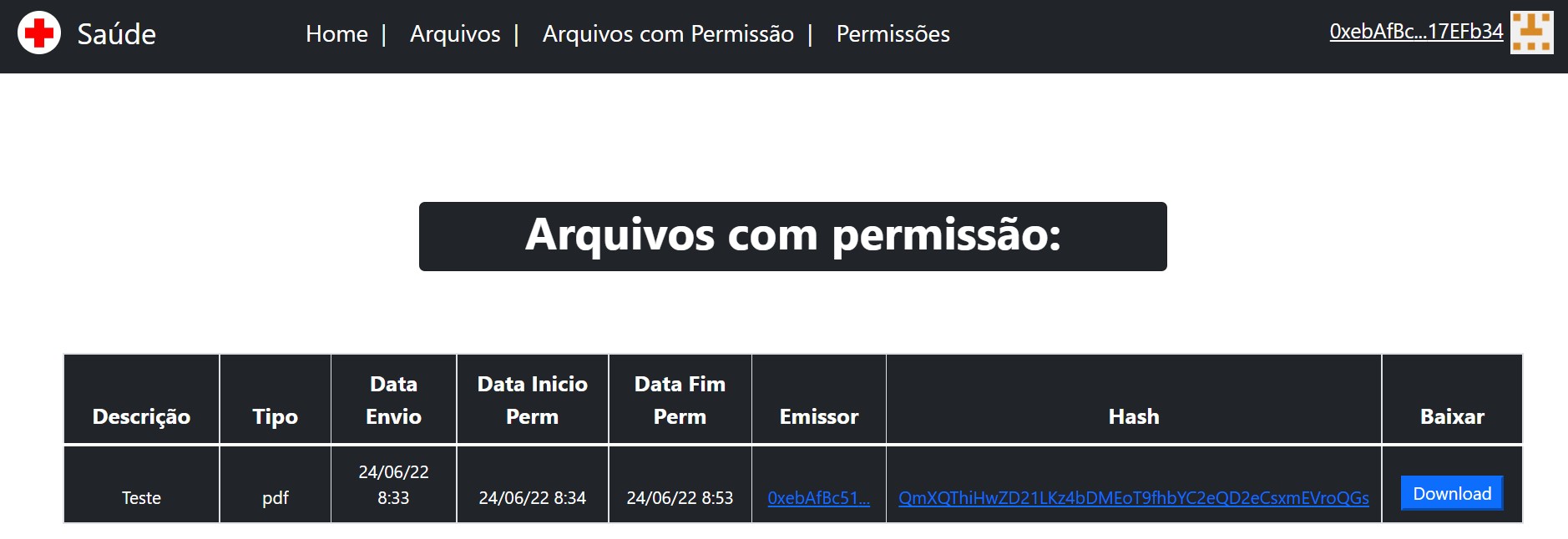}
    \legend{Fonte: Do autor, 2022}
\end{figure}

     Na \autoref{fig:telaPermArq} pode-se obervar a tela responsável por mostrar os arquivos que os outros usuários compartilharam com o usuário que está logado atualmente na DApp, ou seja, os arquivos que o usuário A recebeu a permissão para visualizar e fazer o download, pelo tempo de duração da permissão, de outro usuário B, C entre outros.
    
\chapter{Considerações Finais} \label{consideracoes}
  Este relatório técnico apresentou detalhes da implementação da DApp proposta 
    no trabalho de TCC, que realiza o controle do armazenamento dos dados do 
    usuário de forma descentralizada através da blockchain, o armazenamento dos arquivos no IPFS e da criptografia com os algoritmos RSA ou ECC, em conjunto com AES. O intuito da Dapp é garantir a privacidade dos dados,  além de colocar o controle do gerenciamento desses dados sob responsabilidade do usuário.

    Além de trazer o fundamentos das tecnologias utilizadas, foram abordadas as ferramentas necessárias para o desenvolvimento. A arquitetura da Dapp foi especificada, juntamente com o detalhamento dos seus requisitos funcionais. Foram descritos os mecanismos de comunicação com o MetaMask, a blockchain e o IPFS, bem como detalhados os processos de criptografia com as bibliotecas específicas de cada algoritmo, e apresentada a interface da Dapp para o usuário. 
    
    
    Como trabalhos futuros, pretende-se refatorar o código para obter o melhor desempenho e otimização do código. Além disso, também realizar a implementação da escolha de utilização de um algoritmo de criptografia ou de outro, ou seja, podendo deixar a cargo do usuário a escolha de qual algoritmo  quer utilizar e também deixar a aplicação com suporte para múltiplas chaves criptográficas que o usuário possa possuir.

\postextual  
\bibliography{bibliografia} 
\end{document}